\documentclass[a4paper,11pt]{article}
\usepackage{graphicx} 
\usepackage[english]{babel}
\usepackage{tcolorbox}
\usepackage{enumerate}
\usepackage{mathrsfs}
\usepackage{amsfonts}
\usepackage{amsmath}
\usepackage{ulem}
\usepackage{CJK}
\usepackage{epsfig}
\usepackage{eepic}

\numberwithin{equation}{section}

\usepackage{graphicx}
\usepackage{amssymb}
\usepackage{mdframed} 
\usepackage{indentfirst}
\usepackage{braket}
\usepackage{jheppub}
\newcommand{\pqty}[1]{\left( #1 \right)}
\newcommand{\bqty}[1]{\left[ #1 \right]}

\usepackage{color}
\newcommand{\omits}[1]{}

\arxivnumber{}
\title{\boldmath Entanglement first law for timelike entanglement entropy and linearized Einstein's equation}

\author{Guo-Ying Li$^{1}$} 
\emailAdd{ligy37@mail2.sysu.edu.cn}
\author{Mei-Hui Xiao$^{1}$} 
\emailAdd{xiaomh25@mail2.sysu.edu.cn}
\author{Song He$^{2}$} 
\emailAdd{hesong@nbu.edu.cn}
\author{Jia-Rui Sun$^{1}$} 
\emailAdd{sunjiarui@sysu.edu.cn}

\affiliation{${}^1$School of Physics and Astronomy, Sun Yat-Sen University, Guangzhou 510275, China}
\affiliation{${}^2$Institute of Fundamental Physics and Quantum Technology \& School of Physical Science and Technology,
Ningbo University, Ningbo, Zhejiang 315211, China}

\abstract{We extend the entanglement first law of conformal field theory (CFT) to timelike subregions. Focusing on intervals along the time direction of the boundary CFT, we show that the associated timelike entanglement entropy obeys a first-law–like relation, with an effective entanglement temperature inversely proportional to the temporal size of the interval. By implementing a double Wick rotation, we obtain the exact modular Hamiltonian for a suitable hyperbolic subsystem and use it to formulate the timelike entanglement first law precisely. Our central result is a detailed proof that, in asymptotically Anti-de Sitter spacetime, this timelike entanglement first law is equivalent to linearized Einstein's equations in the bulk: the first law follows from the linearized equations and, conversely, implies them. Our results further reveal the dynamical connections between entanglement and gravity.}

\begin{document}

\maketitle
\flushbottom

\section{Introduction}
Entanglement is a central feature of quantum theory and a powerful probe of quantum many-body systems and gravity. The basic measure is the entanglement entropy, defined as the von Neumann entropy of the reduced density matrix $S_{A} = - \mathrm{tr}_{A}\!\left(\rho_{A} \ln \rho_{A}\right)$, which captures non-local correlations between a region $A$ and its complement $\bar A$ and plays an important role in quantum information and field theory~\cite{Calabrese:2004eu, Levin:2006zz,Kitaev:2005dm,Casini:2022rlv}. In the Anti-de Sitter/Conformal field theory (AdS/CFT) correspondence~\cite{Maldacena:1997re,Gubser:1998bc,Witten:1998qj}, the Ryu--Takayanagi (RT) proposal identifies the entanglement entropy of a spatial subregion $A$ in a $d$-dimensional CFT with the area of a codimension-two bulk minimal surface $\gamma_A$ which is homologous to $A$ in an asymptotically AdS$_{d+1}$ spacetime~\cite{Ryu:2006bv}
\begin{eqnarray}
S_A = \frac{\mathrm{Area}(\gamma_A)}{4 G_N}, \label{RT}
\end{eqnarray}
which is called the RT formula for holographic entanglement entropy (HEE), and has later been extended to a covariant version~\cite{Hubeny:2007xt} and proved by relating entanglement entropy to gravitational entropy~\cite{Lewkowycz:2013nqa,Casini:2011kv,Dong:2016hjy}. This geometric representation of entanglement has motivated studies of bulk reconstruction and the emergence of spacetime in gauge/gravity duality~\cite{Swingle:2009bg,Pastawski:2015qua,Dong:2016eik,Lin:2020ufd,Liang:2025vmx}, and suggested that the dynamics of boundary entanglement encodes the dynamics of the bulk gravity~\cite{VanRaamsdonk:2010pw,Lashkari:2013koa,Swingle:2014uza,Nozaki:2013vta,Bhattacharya:2013bna,deBoer:2015kda,deBoer:2016pqk,Haehl:2017sot}. Building on earlier connections between gravity and thermodynamics, such as black hole mechanics~\cite{Bekenstein:1973ur,Bardeen:1973gs,Hawking:1975vcx} and Jacobson’s derivation of the Einstein equation as an equation of state~\cite{Jacobson:1995ab}, it was shown that in holographic theories where RT surface computes the entanglement entropy, a first law of entanglement based on the modular Hamiltonian~\cite{Blanco:2013joa} implies that small perturbations of the CFT vacuum are dual to spacetime satisfying the Einstein equations linearized around AdS spacetime~\cite{Lashkari:2013koa,Faulkner:2013ica}, with this relation extending beyond linear order perturbations~\cite{Faulkner:2017tkh,Oh:2017pkr}.

A natural generalization of entanglement entropy arises when one replaces the density matrix with a reduced transition matrix between an initial and a final state. The von Neumann entropy of this transition matrix defines the so-called pseudo (entanglement) entropy~\cite{Nakata:2020luh, Mollabashi:2020yie}, which admits a holographic description in terms of minimal surfaces in time-dependent Euclidean AdS geometries. Pseudo entropy has been studied in a variety of systems, including free quantum field theories and spin chains, where it shows nontrivial scaling and saturation behavior~\cite{Mollabashi:2020yie, Mollabashi:2021xsd}, as well as gauge theories and constrained systems, where it is sensitive to gauge structure and topological sectors~\cite{Mukherjee:2022jac}. It also exhibits amplification phenomena, phase transitions, and nontrivial reality as well as pseudo-Hermiticity properties~\cite{Guo:2022jzs, Kanda:2023jyi, Caputa:2024gve}.

Timelike entanglement entropy extends the well-understood concept of spatial entanglement to time-dependent regions. Recent proposals define timelike entanglement entropy in quantum field theories via the analytic continuation of correlation functions, yielding complex results interpreted as pseudo entropy~\cite{Doi:2023zaf, Nakata:2020luh, Mollabashi:2021xsd, Doi:2022iyj}. The relationship between timelike entanglement entropy, pseudo entropy, and spacelike entanglement entropy in the dS/CFT correspondence has been explored in~\cite{Doi:2022iyj, Narayan:2022afv, Narayan:2023zen, Nanda:2025tid}, with recent progress in~\cite{Jiang:2023loq, Jiang:2023ffu, He:2023ubi, Guo:2024lrr, Xu:2024yvf, Milekhin:2025ycm, Anegawa:2024kdj, Zhao:2025zgm, Gong:2025pnu, Guo:2025pru, Nunez:2025gxq, Nunez:2025puk, Jiang:2025pen}. While the RT formula for timelike entanglement entropy remains under development, several proposals for the corresponding RT surfaces exist. In~\cite{Doi:2022iyj, Doi:2023zaf}, timelike entanglement entropy is described by both spacelike RT surfaces and their timelike counterparts, whereas~\cite{Heller:2024whi, Heller:2025kvp, Nunez:2025ppd} suggest that RT surfaces are extremal surfaces in complexified geometry. In $\textrm{AdS}_3$, both approaches yield equivalent results, although distinctions may arise in higher dimensions. This analytic continuation of HEE into timelike subregion offers a novel approach to exploring spacetime emergence. It raises the possibility of a corresponding first-law-like relation for timelike entanglement entropy.


In the present paper, to further reveal the dynamical properties of the timelike entanglement entropy, we will focus on the low-energy excitations of the timelike subregion in CFT on the boundary of an asymptotically AdS spacetime. It is interesting to ask whether a similar entanglement-first law still holds for timelike entanglement entropy, and what its holographic dual is in bulk AdS gravity. By comparing the linearized perturbations of the holographic timelike entanglement entropy (HTEE) and the associated energy of a boundary timelike subregion, we show that a timelike entanglement first law can indeed be introduced via defining an effective entanglement temperature which is proportional to the reciprocal of the time interval. The entanglement temperature is shown to approach a universal value in the small time interval limit, while approaching the Hawking temperature of the dual AdS black hole in the opposite limit. Furthermore, by using the double Wick rotation, we derive the modular Hamiltonian for a hyperbolic subsystem, enabling us to formulate the precise timelike entanglement first law via the variation of the relative entropy. As for the holographic dual of the timelike entanglement first law, it is natural to expect that it corresponds to the bulk linearized Einstein equations containing time evolution. Importantly, we find that a direct calculation of perturbations for HTEE in real space will cause inconsistencies with the timelike entanglement first law. To overcome this problem, we utilize the double Wick rotation to perform the calculations. Finally, we establish the equivalence between the linearized Einstein equations and the timelike entanglement first law, first in the AdS$_4$/CFT$_3$ correspondence and then generalizing to higher-dimensional spacetime. Our results reveal a deep connection between the dynamics of entanglement and gravity, and also provide a timelike analogue of the emergence of spatial directions from entanglement, and suggest a possible mechanism for the emergence of time in the gauge/gravity duality. 

This paper is structured as follows.
In section~\ref{sec-ee1st-et} we first extend the first law of entanglement entropy to timelike subregion in CFT on the boundary of asymptotically AdS$_{d+1}$ spacetime and derive the corresponding entanglement temperature. Then employing double Wick rotation, we derive the modular Hamiltonian and timelike entanglement first law for hyperbolic subsystems. In section~\ref{sec-Eineq-ee1stlaw} we prove the equivalence between linearized Einstein's equations and the timelike entanglement first law, generalizing from the AdS$_{4}$/CFT$_{3}$ correspondence to higher dimensions. In section~\ref{sec-conclu}, we present the conclusions and discuss future related directions for investigation. The appendix provides details on the main context.

\textbf{ Note:} At the completion stage of the present paper, we notice that the paper \cite{Fujiki:2025rtx} appeared on the Arxiv, which studied a similar topic as in our paper. The authors showed that the first law of holographic pseudo-entropy in dS$_3$ spacetime is equivalent to the perturbative Einstein equation, thereby highlighting the emergence of the time direction in dS$_3$ spacetime. 

\section{Timelike entanglement first law and entanglement temperature}\label{sec-ee1st-et}
When adding low-energy perturbations to a CFT, the variations of entanglement entropy $\Delta S$ and the energy $\Delta E$ of a small subsystem obey an equation that is analogous to the first law of thermodynamics, called the entanglement first law~\cite{Bhattacharya:2012mi}
\begin{eqnarray}\label{ee1stlaw}
    T_{\rm ent}\cdot\Delta S=\Delta E,
\end{eqnarray}
where $T_{\rm ent}=c\cdot l^{-1}$ is an effective temperature called the entanglement temperature, $l$ is the size of the subsystem and $c$ is an order one constant. For example, $c=\frac{d+1}{2\pi}$ when the
subsystem is a round sphere. This relation was first derived by considering low-energy excitations of the HEE of CFT$_2$, and later, eq.(\ref{ee1stlaw}) was obtained from linear variation of the relative entropy and the modular Hamiltonian~\cite{He:2014lfa}, and it was shown that the entanglement first law becomes an inequality at second order variations~\cite{Blanco:2013joa}. The entanglement first law provides a
universal relationship between the energy and the amount of quantum
information. In this section, we will similarly proceed from holographic
calculations and consider thermal excitations for a small subsystem.
We aim to demonstrate that an analogous property, resembling the first law of thermodynamics, also holds for timelike entanglement entropy. In this case, the effective temperature remains proportional to the inverse of the subsystem size, where the subsystem size is now a time interval.

\subsection{Strip subsystem: $\left|t\right|<\frac{\Delta t}{2}$}
Our first choice of subsystem $A$ is a strip living in $d$-dimensional
Minkowski spacetime located on the regularized ($z=\varepsilon\ll1$)
boundary of the asymptotically AdS spacetime
\begin{equation}
ds^{2}=\frac{R^2_{\rm AdS}}{z^{2}}\left(-f(z)dt^{2}+\frac{dz^{2}}{f(z)}+d\mathbf{x}^{2}\right),\quad \mathbf{x}\in\mathbb{R}^{d-1}.\label{bulk metric}
\end{equation}

Without loss of generality, throughout this work, the radius $R_{\rm AdS}$
is set to unity for simplicity. The choice $f(z)=1$ corresponds to the pure AdS space dual to the vacuum of the
boundary CFT, while $f(z)=1-\frac{z^{d}}{z_{h}^{d}}$ corresponds to a AdS black hole dual to the boundary thermal CFT. 

It was proposed in~\cite{Heller:2024whi} that the bulk carriers
of the HTEE are codimension-two
extremal surfaces $\gamma_{\mathcal{T}}$ anchored on a timelike boundary
subregion $\mathcal{T}$ and in general extending in a complexified
bulk geometry. The HTEE is then proportional to the area of $\gamma_{\mathcal{T}}$,
\begin{equation}
S_{A}=\frac{{\rm Area}(\gamma_{\mathcal{T}})}{4G_{N}}.
\end{equation}

The strip is defined as $\mathcal{T}\equiv\left\{ (t,\mathbf{x})|-\frac{\Delta t}{2}<t<\frac{\Delta t}{2},-L/2<x_{1,2,\cdot\cdot\cdot,d-2}<L/2,x_{d-1}=0\right\} $, where $L$ is taken to be infinite. By symmetry, we can parameterize the
minimal surface by $t=t(z)$. Then its area is computed as
\begin{equation}
A=\int dz\,\frac{1}{z^{d-1}}\sqrt{\frac{1}{f(z)}-f(z)t'(z)^{2}}.\label{eq:area func}
\end{equation}

In principle, we could extremize the above expression, but this would
require a numerical evaluation and the HTEE for the strip $\mathcal{T}$
is only known in the vacuum~\cite{Heller:2024whi}. To make progress analytically,
we will carry out a perturbative calculation for a small interval $\Delta t$
and low temperatures, in which we consider the limit $R_{\rm AdS}/z_{h}\ll1$.
In this case, the minimal surface is only probing the asymptotic region
of the black brane geometry, and so the solution deviates only slightly
from the vacuum solution, i.e., $t(z)=t_{0}(z)+\delta t(z)$. The
deviation will not modify the result at first order in our perturbative
calculation, and the leading order change in the entropy comes from
evaluating eq.(\ref{eq:area func}) with $t=t_{0}(z)$ \cite{Blanco:2013joa}.

Thus we first calculate the vacuum solution $t_{0}(z)$ for $f(z)=1$ in
eq.(\ref{eq:area func}). Since the bulk metric eq.(\ref{bulk metric}) does
not depend on time, the extremal surfaces $\gamma_{\mathcal{T}}$
has an associated conserved quantity, $p$, such that the Euler–Lagrange
equations stemming from eq.(\ref{eq:area func}) can be reduced to the
first-order form
\begin{equation}
{\footnotesize p:=\frac{\partial\mathcal{L}}{\partial t'}=\frac{-t'(z)}{z^{d-1}\sqrt{1-t'(z)^{2}}},\quad t'(z)=\frac{pz^{d-1}}{\sqrt{1+p^{2}z^{2d-2}}}}.\label{Euler=002013Lagrange}
\end{equation}

From eq.(\ref{Euler=002013Lagrange}) it is immediate to see that the locus
$z=z_{t}$ where $1+p^{2}z^{2d-2}=0$ corresponds to a tip of $\gamma_{\mathcal{T}}$
where $t_{0}(z)$ has a branch-point singularity. For the vacuum state,
the solution of eq.(\ref{Euler=002013Lagrange}) subject to the boundary
conditions $t_{\pm}(z=\epsilon\to0)=\pm\frac{\Delta t}{2}$ is given
by
\begin{equation}
t_{\pm}(z)=\pm\frac{\Delta t}{2}\pm\frac{iz_{t}}{d}\left(\frac{z}{z_{t}}\right)^{d}\cdot{}_{2}F_{1}\left(\frac{1}{2},\frac{d}{2(d-1)},\frac{3d-2}{2(d-1)},\left(\frac{z}{z_{t}}\right)^{2d-2}\right),
\end{equation}

where $z_{t}$ is the tip of the extremal surface
\begin{equation}
 z_{t}=\frac{i\,\Gamma\left(\frac{1}{2(d-1)}\right)}{2\sqrt{\pi}\Gamma\left(\frac{d}{2(d-1)}\right)}\cdot\Delta t.
\end{equation}

Expanding eq.(\ref{eq:area func}) to leading order in $1/z_{h}^{d}$ yields
\begin{equation}
{\scriptsize \Delta A=\left.2L^{d-2}\int_{0}^{z_{t}}dz\,\frac{1+t'(z)^{2}}{2z^{-1}z_{h}^{d}\sqrt{1-t'(z)^{2}}}\right|_{t=t_{0}(z)}}.
\end{equation}
The entanglement entropy becomes
\begin{equation}
\Delta S_{A}=-\frac{1}{16G_{N}}L^{d-2}\cdot\Delta t^{2}\cdot\frac{1}{z_{h}^{d}}\cdot\underset{C}{\underbrace{\frac{\Gamma\left(\frac{1}{2(d-1)}\right)^{2}}{2(d-1)\pi\Gamma\left(\frac{d}{2(d-1)}\right)^{2}}\cdot\left[\frac{\Gamma\left(\frac{1}{d-1}\right)\Gamma\left(\frac{1}{2}\right)}{\Gamma\left(\frac{1}{d-1}+\frac{1}{2}\right)}-2\cdot\frac{\Gamma\left(\frac{d}{d-1}\right)\Gamma\left(\frac{1}{2}\right)}{\Gamma\left(\frac{d}{d-1}+\frac{1}{2}\right)}\right]}.}\label{deltaSstrip}
\end{equation}

Specifically $d=2$, $C=-\frac{1}{3}; d=3,C=0;d\geq4,C<0.$ 

The deviation of the entanglement entropy is only related to the metric perturbation, while the derivation of the subsystem's energy depends on the stress tensor. To find the relation between the boundary stress tensor and this metric perturbation, we apply a coordinate transformation to eliminate the perturbation in the $z$-component. This allows us to use the Fefferman-Graham (FG) expansion \cite{Fefferman:2007rka} to express the stress tensor in terms of the perturbed metric~\cite{Blanco:2013joa}.

Consider the transformation defined by
\begin{equation}
\frac{d\xi}{\xi} = \frac{dz}{z f^{1/2}(z)}.
\end{equation}
Integrating this, we find the relation $z = \xi \left(1 - \frac{\xi^d}{2 d z_h^d} + \mathcal{O}\left(\frac{\xi^{2d}}{z_h^{2d}}\right)\right)$. We can now rewrite the perturbed metric eq.(\ref{bulk metric}) in the new coordinates as
\begin{align}
d\hat{s}^2 = \frac{1}{\xi^2}\left[-\left(1 + \frac{1-d}{d}\frac{\xi^d}{z_h^d}\right)dt^2 + d\xi^2 + \left(1 + \frac{\xi^d}{d z_h^d}\right)\sum_{i=1}^{d-1}dx_i^2\right].
\end{align}
This transformation maps the thermal perturbation background to a metric perturbation that no longer has a direct thermodynamic interpretation. However, this change does not affect the boundary stress tensor. The stress tensor is defined on the asymptotic AdS boundary $\xi \to 0$ (equivalent to $z \to 0$), and the transformation is purely radial, meaning it leaves the boundary geometry unchanged \cite{Skenderis:2000in}. For simplicity, we assume that the stress tensor is constant (which means that we restrict the stress tensor on the boundary), namely,
\begin{equation}
T_{tt}=\frac{\left(d-1\right)}{16\pi G_{N}}\left(\frac{1}{z_{h}}\right)^{d}.
\end{equation}
Thus, the increased amount of energy in the subsystem $A$ is given by
\begin{equation}\label{deltaEstrip}
\Delta E_{A}=\int\Delta T_{tt}\cdot dx^{d-1}=\frac{1}{16\pi G_{N}}L^{d-2}\frac{(d-1)}{z_{h}^{d}}\cdot\Delta t.
\end{equation}

Therefore, comparing eq.(\ref{deltaSstrip}) with eq.(\ref{deltaEstrip}), we find the following relation,
\begin{equation}
T_{\rm ent}=\frac{\Delta E_{A}}{\Delta S_A}=-\frac{(d-1)}{C\pi}\cdot\left(\Delta t\right)^{-1},\label{strip T}
\end{equation}
which shows that a timelike entanglement first law $T_{\rm ent} \Delta S_A=\Delta E_A$ can be introduced by defining an entanglement temperature $T_{\rm ent}$ which is inversely proportional to the size of the time interval $\Delta t$.

\subsection{Hyperbolic subsystem: $-t^{2}+\sum_{i=1}^{d-2}x_i<-T_{0}^{2}$}
As another example, consider the case where the subsystem $A$ is defined by the hyperbolic region $-t^2+\sum_{i=1}^{d-2}x_i^2\leq -T_0^2$. We first compute the timelike entanglement entropy by applying the standard Wick rotation $t_E=it, R=iT_0$. After obtaining the entanglement entropy for $A$ via this Wick rotation in the context of Euclidean AdS$_{d+1}$ (EAdS$_{d+1}$) spacetime, we identify the geometric configuration that reproduces the Wick-rotated result \cite{Doi:2023zaf}. The metric of pure Poincar\'{e}  EAdS$_{d+1}$ spacetime is 
\begin{align}
    ds^2&=\frac{1}{z^2}\left(g_{zz}dz^2+g_{\mu\nu}dx^{\mu}dx^{\nu}\right),\notag\\
    &=\frac{1}{z^2}\bigg(dt_E^2+dz^2+\sum_{i=1}^{d-1}dx_i^2\bigg),
\end{align}
where $\mu,\nu=0,1,\dots,d-1\notag$. The induced metric over the extremal surface $z^2+r^2=R^2$ \cite{Doi:2023zaf} of the subregion $-R<r<R$, where $r^2=t_E^2+\sum_{i=1}^{d-2}x_i^2, R=iT_0$, is
\begin{align}
    H_{ij}&=\frac{1}{z^2}\big(g_{ij}+g_{zz}\partial_iz \partial_jz\big),\quad {\rm with}\quad i,j=0,1,\dots,d-2.
\end{align}
The corresponding area is 
\begin{align}
    A&=\int d^{d-1}x \sqrt{H},\notag\\
    &=\int d^{d-1}x \frac{1}{z^{d-1}}\sqrt{{\rm det}g_{ij}}\sqrt{1+g_{zz}g^{ij}\partial_i z\partial _j z}.
\end{align}
Now, consider a low-energy thermal perturbation, the metric becomes
\begin{align}
    d\hat{s}^2&=\frac{1}{z^2}\left(\hat{g}_{zz}dz^2+\hat{g}_{\mu\nu}dx^\mu dx^\nu\right),\notag\\
    &=
    \frac{1}{z^2}\bigg(f(z)dt_E^2+\frac{dz^2}{f(z)}+\sum_{i=1}^{d-1}dx_i^2\bigg),\label{perturbation metric}
\end{align}
where $\mu,\nu=0,1,\dots,d-1 \notag$. Then the first order perturbation gives 
\begin{align}
    \delta g_{t_E t_E}&=-\frac{z^d}{z_h^d}, \\
    \delta g_{zz}&=\frac{z^d}{z_h^d}.
\end{align}
Similarly, to obtain an analytical solution, we employ a perturbative approach, i.e., the shape of the extremal surface deviates only slightly from that in pure Poincar\'{e} EAdS$_{d+1}$ as $z'(r) = z(r) + \delta z(r)$, where $z(r)$ describes the original extremal surface. Since $z(r)$ already extremizes the area, the deviation $\delta z(r)$ will not modify the result at first order in our perturbative calculation. This means the first-order change in the area functional arises solely from changes in the metric itself.
\begin{align}
    \delta A
    &=\int d^{d-2} x dt_E \frac{1}{z^{d-1}}\frac{1}{2}\sqrt{g}\sqrt{1+g_{zz}g^{ij}\partial_i z\partial_j z}\bigg(g^{ij}\delta g_{ij}+\frac{\delta g_{zz}g^{ij}\partial_i z\partial_j z+g_{zz}\delta g^{ij}\partial_i z\partial_j z}{1+g_{zz}g^{ij}\partial_i z\partial_j z}\bigg),\notag\\
    &=\frac{1}{2  z_h^d}\int d^{d-2}xdt_E\frac{(r^2+t_E^2-R^2)}{R}.
\end{align}
The deviation of entanglement entropy is 
\begin{align}
    \Delta S_E&=2\pi \frac{\delta A}{l_p^{d-1}},\notag\\
    &=-\frac{\pi }{ z_h^d l_p^{d-1}}\frac{R^{d}\Omega_{d-2}}{(d^2-1)}.\label{sh}
\end{align}

The corresponding perturbation of energy related to this subsystem is
\begin{align}
    \Delta E_{E} &=\Delta T_{t_E t_E}\int_{t_E^2+\sum_{i=1}^{d-2}x_i^2\leq R^2}(-i) dt_Edx^{d-2},\notag\\
    &=i \Delta T_{tt}\Omega_{d-2}R^{d-1}\frac{1}{(d-1)},\notag\\
    &=i\frac{1}{2}\frac{1}{l_p^{d-1}z_h^d}\frac{\Omega_{d-2}R^{d-1}}{(d-1)},
\end{align}
which gives
\begin{align}
    T_E=\frac{\Delta E_{E}}{\Delta S_E}=-i\frac{(d^2-1)}{2\pi}R^{-1}.
\end{align}
Finally, performing the Wick rotation back by substituting $R=iT_0$, the entanglement temperature is
\begin{align}
    T_{\rm ent}&=-\frac{(d^2-1)}{2\pi}T_0^{-1}.
\end{align}

\subsection{Comparison with CFT at finite $T$}

\begin{figure}[h]
\includegraphics[scale=0.5]{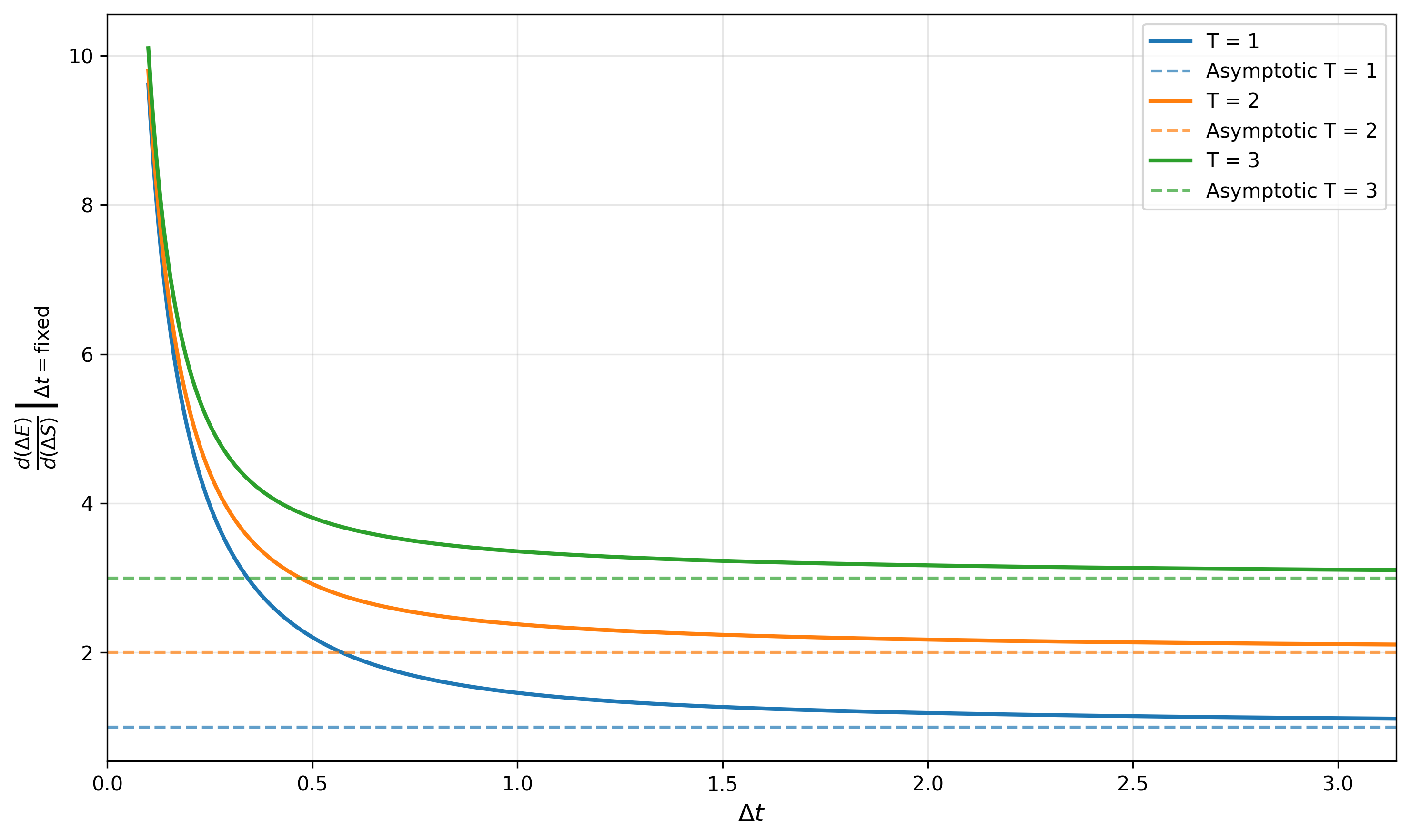}\caption{The entanglement temperature $T_{\rm ent}$=$\frac{d(\Delta E)}{d(\Delta S)}|_{\Delta t=\textrm{fixed}}$ of CFT$_2$ as
a function of $\Delta t$ based on the holographic calculation: the bulk dual is the BTZ black hole at temperature $T$, where we have set $T=1,2,3$.}
\label{field et}
\end{figure}

As has been discussed in \cite{Bhattacharya:2012mi} that from the field theory side, the entanglement temperature $T_{\rm ent}$ defined from the entanglement first law in eq.(\ref{ee1stlaw}) has a universal expression when the spatial size of the subsystem becomes very small, which corresponds to the large $N$ limit in dual strongly coupled field theory. Now, for the timelike entanglement entropy case, we expect a similar relation to hold, namely, 
\begin{align}
    T_{\rm ent}=\text{constant}\cdot t^{-1}, \label{ent T}
\end{align}
is satisfied if the size of subsystem  $\Delta t$ is sufficiently small such that
\begin{equation}
T_{tt}\cdot\Delta t^{d}\ll R_{\textrm{AdS}}^{d-1}/G_{N}\sim\mathcal{O}(N^{2}).\label{condition}
\end{equation}
\omits{ (\ref{first law}) is the universal statement that
the amount of information included in a small subsystem $A$ is proportional
to the energy included in $A$. The AdS/CFT correspondence predicts that the constant
c in \ref{ent T} is universal when the shape of the subsystem
$A$ fixed. Since the source of the excitation energy is arbitrary, it
can be a temperature increase or the creation of massive objects at zero
temperature.} The limit in eq.(\ref{condition}) is crucial, and if it is not satisfied, then $T_{\rm ent}$ as defined by eq.(\ref{ee1stlaw}) is
no longer universal and will depend on the details of the excitations.
On the gravity side, the result depends on the details of the infrared
region. For instance, consider the AdS black hole at Hawking temperature $T$, $T_{\rm ent}$
defined by eq.(\ref{ee1stlaw}) approaches eq.(\ref{ent T}) in the $\Delta t\rightarrow0$ limit,
while approaches to $T$ up to a constant factor in the opposite limit $\Delta t\to\infty$. To show this, consider a two-dimensional CFT on a cylinder with finite temperature, whose gravity dual
 is the BTZ black hole
\begin{equation}
ds^{2}=-(r^{2}-r_{+}^{2})dt^{2}+\frac{dr^{2}}{(r^{2}-r_{+}^{2})}+r^{2}d\phi^{2},
\end{equation}
where $r_{+}=\frac{2\pi}{\beta}$ is the radius of the horizon. Choosing $A$ as a timelike interval with length $\Delta t$, the timelike entanglement entropy can be obtained from the geodesic length,
\begin{equation}
S_{\mathcal{T}}=\frac{c}{3}\log{\left(\frac{\beta}{\pi\epsilon}\sinh\left(\frac{\pi\Delta t}{\beta}\right)\right)}+\frac{c\pi}{6}i.
\end{equation}
The entanglement entropy of the ground state is given by \cite{Doi:2023zaf}
\begin{align}
    S=\frac{c}{3}\log{\frac{\Delta t}{\epsilon}}+\frac{c \pi}{6}i.
\end{align}
Thus, the variation of the entanglement entropy is
\begin{align}
    \Delta S&=S_{\mathcal{T}}-S,\notag\\
    &=\frac{c}{3}\log{\left(\frac{\beta}{\pi \Delta t}\sinh{\left(\frac{\pi \Delta t}{\beta}\right)}\right)},
\end{align}
which is ultraviolet (UV) divergence-free.
Since the related energy is $\Delta E=\frac{c\pi}{6\beta^2}\Delta t$, then the expression of the entanglement temperature is
\begin{align}
    T_{\rm ent}&=\frac{\Delta E}{\Delta S},\notag\\
    &=\frac{r_+^2\Delta t}{8\pi \log{\left(\frac{2}{r_+\Delta t}\sinh{\left(\frac{r_+\Delta t}{2}\right)}\right)}},
\end{align}
 Considering the UV limit $\Delta t\to0$,
\begin{align}\label{Tuvcft2}
    T_{\rm ent}&=\frac{3}{\pi}\frac{1}{\Delta t},
\end{align}
the entanglement temperature obeys eq.( \ref{ent T}). However, if we take the infrared (IR) limit $\Delta t\to \infty$, the entanglement temperature is not the exact Hawking temperature
\begin{align}
    T_{\rm ent}&=\frac{r_+}{4\pi}=\frac{T}{2}.
\end{align}
To eliminate the constant factor, we need to redefine the first law of entanglement as
\begin{align}
T_{\rm ent}&=\frac{d\Delta E}{d\Delta S},\label{redefine ent T} \\
 &=\frac{\frac{\Delta t}{4\pi}r_{+}}{-\frac{1}{r_{+}}+\frac{\Delta t}{2}\coth\left(\frac{\Delta tr_{+}}{2}\right)} .
\end{align}
Then, considering the UV behavior,
\begin{align}
\frac{d\Delta E}{d\Delta S} 
& =\frac{\frac{\Delta t}{4\pi}r_{+}}{-\frac{1}{r_{+}}+\frac{\Delta t}{2}\left(\frac{2}{\Delta tr_{+}}+\frac{\Delta tr_{+}}{6}+O(\Delta t^{3})\right)},\nonumber \\
 & =\frac{\frac{\Delta t}{4\pi}r_{+}}{\frac{(\Delta t)^{2}r_{+}}{12}+O((\Delta t)^{3})}\rightarrow\frac{3}{\pi\Delta t},
\end{align}
the entanglement temperature is the same as the original expression in eq.(\ref{Tuvcft2}).
Nevertheless, if we take the IR limit, i.e., $\Delta t\to\infty$,
\begin{align}\label{ee1stder}
\frac{d\Delta E}{d\Delta S} & =\frac{\frac{\Delta t}{4\pi}r_{+}}{-\frac{1}{r_{+}}+\frac{\Delta t}{2}\big(1+2e^{-\Delta tr_{+}}+O(e^{-2\Delta tr_{+}})\big)}\rightarrow\frac{r_{+}}{2\pi}=T.
\end{align}
This definition yields the exact Hawking temperature. 
The evolution of the entanglement temperature along the renormalization-group (RG) flow is illustrated in Fig. \ref {field et}.

\subsection{More on the timelike entanglement first law}\label{argue}
So far, we have calculated the entanglement temperature between the ground state and the thermally excited state for different shapes of timelike subsystems. To verify whether the entanglement temperature we considered above approaches the Hawking temperature in the IR limit, we studied a special case in $d=2$. The result showed that, to obtain the Hawking temperature exactly, the entanglement first law must be redefined as in eq.(\ref{redefine ent T}). Otherwise, the final result differs by a constant factor. In the IR limit, this definition is well justified, since it reduces to the first law of black hole thermodynamics, given that the renormalized entanglement entropy becomes thermal entropy in the IR region \cite{Kim:2016jwu}. Nevertheless, the physical meaning of the definition in eq.(\ref{redefine ent T}) in the UV limit becomes ambiguous, as it no longer resembles the first law of black hole due to the appearance of two derivatives. 

To avoid the ambiguity, a modified definition of the entanglement first law was used in \cite{Saha:2019ado, Kim:2016jwu}, which employed the integral form of the first law of black hole; the entanglement first law can be rewritten as
\begin{align}\label{modee1stlaw}
    \Delta E=\frac{(d-1)}{d}T_{\rm ent} \Delta S.
\end{align}
This approach yields the exact Hawking temperature and resolves ambiguity in the UV limit. However, this constant factor, $\frac{d-1}{d}$, is not universal:  different black hole excitations produce different constant factors, as discussed in  \cite{Saha:2020fon}. Moreover, both the definition in eq.(\ref{redefine ent T}) and eq.(\ref{modee1stlaw}) rely explicitly on the entanglement temperature.

To overcome these problems, we adopt the definition proposed in \cite{Blanco:2013joa}, which is based on the non-negativity of the relative entropy. This ensures that the definition remains valid along the RG flow. The variation of the relative entropy $S(\rho_1|\rho_0)={\rm tr}(\rho_1\ln\rho_1)-{\rm tr}(\rho_1\ln\rho_0)$ gives
\begin{align}\label{relee1stlaw}
    \Delta S\leq \Delta\langle H\rangle,
\end{align}
where $\rho_0$ and $\rho_1$ are the reduced density matrices of two states of a given subsystem. $H$ is the modular Hamiltonian associated with the subsystem defined by
\begin{eqnarray}
    \rho=\frac{e^{-H}}{{\rm tr}e^{-H}},
\end{eqnarray}
and $\langle H\rangle={\rm tr}(\rho H)$ is the average value. Since the relative entropy quantifies the difference between two states, this formulation can be applied to arbitrary excitations. The inequality becomes saturated when the distinction between the two states is small. In the following discussion, we assume the subsystem is sufficiently small to analyze the timelike entanglement first law.

Since the form of the modular Hamiltonian (see appendix \ref{modularh}) for the specific geometric region (the hyperbolic subregion) is known, we can verify that the first-order term of the relative entropy indeed vanishes, thereby confirming the entanglement first law in this case.

After double Wick rotation, the modular Hamiltonian of the hyperbolic subsystem is
\begin{align}
\Delta \langle H \rangle
&=-\frac{\pi }{2l_p^{d-1}z_h^d} \int d^{d-1}x\frac{(R^2-r^2)}{R},\notag\\
&=-\frac{\pi }{l_p^{d-1}z_h^d}\Omega_{d-2}\frac{R^d}{(d^2-1)} .\label{mh in double wick}
\end{align}
Because eq.(\ref{sh}) is independent of the coordinate $x_{d-1}$,  performing a Wick rotation $x_{d-1}\to i x_{d-1}$ leaves the final result unchanged. Comparing with eq.(\ref{mh in double wick}) then yields 
\begin{align}
    \Delta S=\Delta \langle H \rangle,
\end{align}
which implies that the entanglement first law also holds for timelike entanglement entropy in the hyperbolic subsystem case.

\section{Linearized Einstein's equation and timelike entanglement first law}\label{sec-Eineq-ee1stlaw}
In this section, we continue our study of the hyperbolic subregion defined by $t^2-\sum_{i=1}^{d-2}x_i^2\geq T_0^2$ and choose $x_{d-1}=0$. Our goal is to verify whether the entanglement first law can be derived from the linearized Einstein's equation, or conversely, if the linearized Einstein's equation can be derived from the entanglement first law. To explicitly see this relation, we begin with the lower-dimensional case of $d=3$ (the $d=2$ case is relevant only to the traceless condition of the perturbation metric).
 To make progress more tractable, we assume that the perturbation satisfies the FG expansion. This condition allows us to find a direct relation between the boundary stress tensor and the bulk perturbative metric. Because the modular Hamiltonian is ill-defined in Euclidean spacetime, and its calculation in hyperbolic spacetime leads to an IR divergence caused by the infinite volume of the hyperbolic region, we must use a regulated definition. Moreover, we find that when calculating the perturbation of the timelike entanglement entropy in real space, the perturbed area cannot be regarded as the union of spacelike and timelike areas. This interpretation is inconsistent with the entanglement first law, as we have shown in Appendix \ref{real}.\omits{~\footnote{The same issue also arises in \cite{Fujiki:2025rtx}, where such a decomposition is shown to be inappropriate.}}
 Therefore, unless otherwise stated, we assume that both the entanglement entropy and the modular Hamiltonian are computed using the holographic formula in the spacetime obtained after the double Wick rotation, i.e.,
\begin{eqnarray}
  t\rightarrow -it_E\quad {\rm and}\quad x_{d-1}\to ix_{d-1},
\end{eqnarray}
the hyperbolic subsystem then becomes $t_E^2+\sum_{i=1}^{d-2}x_i^2\leq R^2$, where $ R=iT_0$. The metric of the perturbed geometry around pure AdS$_{d+1}$ spacetime after double Wick rotation is
\begin{align}
ds^2=\frac{1}{z^2}\left(dz^2+(g_{\mu\nu}+\delta g_{\mu\nu})dx^\mu dx^\nu\right), \quad \mu,\nu=0,1,\dots,d-1.\label{metric for FG expansion}
\end{align}
Here, $g_{\mu\nu}=\text{diag}\{1,\dots,1,-1\}$ is the background metric, and $\delta g_{\mu\nu}=z^d h_{\mu\nu}$ is the perturbation. The area of extremal surface changes in this way
\begin{align}\label{deltaA}
    \delta A=\frac{1}{2 R}\int d^{d-1}x \frac{1}{z^d}\big(R^2g^{ij}-x^i x^j\big)\delta g_{ij},\quad i,j=0,\dots,d-2.
\end{align}
Using the FG expansion, the perturbation of the bulk metric takes the form \cite{Blanco:2013joa}
\begin{align}
\delta g_{\mu\nu}(z,x,t)=\frac{2l_p^{d-1}}{d}\sum_{n=0}^\infty z^{2n} T^{(n)}_{\mu\nu}(x,t), \label{metric and stress tensor}
\end{align}
where $l_p^{d-1}$ relates the $d+1$-dimensional Newtonian constant. For brevity, we omit the superscript $E$, which indicates that the stress tensor is considered in the double Wick rotated spacetime. The formula allows us to describe the stress tensor near the boundary. According to Einstein's equations, all coefficients $T^{(n)}_{\mu\nu}$ can be expressed in terms of the expectation value of the boundary stress tensor $T^{(0)}_{\mu\nu}$. In this work, we only consider terms up to linear order in $T^{(0)}_{\mu\nu}$. We will then generalize our discussion to arbitrary dimensions and show that, even in the original AdS$_{d+1}$ spacetime, namely, $ds^2 = \frac{1}{z^2}(-dt^2 + dz^2 + \sum_{i=1}^{d-1} dx_i^2)$, in which IR divergences are present, we can still obtain these results.

\subsection{The case of $d=3$}
The Einstein's equations of the metric in eq.(\ref{metric for FG expansion}) up to linearized order in $4$-dimension are \cite{Lashkari:2013koa} 
\begin{align}
    h^\mu_\mu=0,\quad \partial_\mu h^{\mu\nu}=0,\quad \frac{1}{z^4}\partial_z(z^4\partial_zh_{\mu\nu})+\partial^2h_{\mu\nu}=0,\quad {\rm with}\quad \mu,\nu=0,1,2. \label{eq for h}
\end{align}
and according to eq.(\ref{metric and stress tensor}), Einstein's equations can be reformulated in terms of stress tensors as
\begin{align}
    T^{(n)\mu}_\mu=0,\quad \partial_\mu T^{(n)\mu\nu}=0,\quad T^{(n)}_{\mu\nu}=-\frac{\partial^{2}T^{(n-1)}_{\mu\nu}}{(2n)(2n+3)}.\label{eq}
\end{align}
\subsubsection{Proof of linearized Einstein's equations will lead to the entanglement first law in AdS$_4$/CFT$_3$}
To obtain the entanglement entropy from linearized Einstein's equations, we assume that our subsystem is centered at any point $(t_{E0},x_0)$ on the boundary. The linearized variation of the modular Hamiltonian is 
\begin{align}
    \Delta \langle H \rangle=&\frac{\pi}{R}\sum_{m_x,m_t=0}^{\infty}\frac{1}{(2m_x)!}\frac{1}{(2m_t)!}\partial_{x_0}^{2m_x}\partial_{t_{E0}}^{2m_t} T^{(0)}_{yy}(x_0,t_{E0},y)\int_{t_E^2+x^2\leq R^2}dt_E dx\notag \\
    &\times\big(R^2-t_E^2-x^2\big)x^{2m_x}t_E^{2m_t}, \label{mh in 3}
\end{align}
where we have used the Taylor expansion to express the stress tensor at the point $(t_{E0},x_0)$. Due to symmetry, the odd term vanishes. 
With the help of eqs.(\ref{deltaA},\ref{metric and stress tensor}), we can express the derivation of the entanglement entropy as a series sum of the stress tensor $T^{(n)}_{\mu\nu}$ located a short distance from the boundary
\begin{align}
    \Delta S
    =&\frac{2\pi}{3 R} \bigg(\sum_{n,m_x,m_t=0}^\infty \frac{1}{(2m_x)!}\frac{1}{(2m_t)!}\partial_{x_0}^{2m_x} \partial_{t_{E0}}^{2m_t}\int dxdt_Ex^{2m_x}t^{2m_t}\big(R^2-t_E^2-x^2\big)^n \notag\\
   & \times\Big(T^{(n)}_{t_Et_E}(t_{E0},x_0,y)(R^2-t_E^2)+T_{xx}^{(n)}(t_{E0},x_0,y)(R^2-x^2)\Big)\notag\\
   &-2\sum_{n,m_x,m_t=0}^\infty\frac{1}{(2m_x+1)!}\frac{1}{(2m_t+1)!}\partial_{x_0}^{2m_x+1}\partial_{t_{E0}}^{2m_t+1}\notag \\&\times\int dx dt_E x^{2m_x+2}t_E^{2m_t+2}(R^2-t_E^2-x^2)^nT^{(n)}_{xt_E}(t_{E0},x_0,y)\Big).
\end{align}
The result can be expressed in powers of $R$ through basic integration as
\begin{align}
    \int_{t_E^2+x^2\leq R^2} dt_E dt \big(R^2-t^2_E-x^2\big)^nx^{2m_x}t_E^{2m_t}=R^{2+2m_x+2m_t+2n}I_{m_x,m_t,n},
\end{align}
where
\begin{align}
I_{m_x,m_t,n}=\frac{\Gamma(\frac{1}{2}+m_t)\Gamma(\frac{1}{2}+m_x)\Gamma(1+n)}{\Gamma(2+m_t+m_x+n)}.\label{integration}
\end{align}
And applying eq.(\ref{integration}), we have
\begin{align}
    \Delta S =&\frac{2\pi}{3R}\sum_{m_x,m_t,n=0}^\infty R^{4+2m_x+2m_t+2n}\bigg(\frac{1}{(2m_x)!}\frac{1}{(2m_t)!}\partial_{x_0}^{2m_x}\partial_{t_{E0}}^{2m_{t}}T^{(n)}_{t_Et_E}(t_{E0},x_0,y)\big(I_{m_x,m_t,n}-I_{m_x,m_t+1,n}\big)\notag\\
   &+\frac{1}{(2m_x)!}\frac{1}{(2m_t)!}\partial_{x_0}^{2m_x}\partial_{t_{E0}}^{2m_{t}}T^{(n)}_{xx}(t_{E0},x_0,y)\big(I_{m_x,m_t,n}-I_{m_x+1,m_t,n}\big)\notag\\
   &-2\frac{1}{(2m_x+1)!}\frac{1}{(2m_t+1)!}\partial_{x_0}^{2m_x+1}\partial_{t_{E0}}^{2m_{t}+1}T^{(n)}_{xt_E} R^2I_{m_x+1,m_t+1,n}\bigg).\label{series of S}
\end{align}
Note that Einstein's eq.(\ref{eq}) can be rewritten as 
\begin{align}
    2\partial_{t_E} \partial_x T^{(n)}_{xt_E}=\partial_x^2T^{(n)}_{t_Et_E}+\partial_{t_E}^2T^{(n)}_{xx}+(2n+2)(2n+5)\Big(T^{(n+1)}_{t_Et_E}+T^{(n+1)}_{xx}\Big).
\end{align}
By applying this formula, eq.(\ref{series of S}) reduces to
\begin{align}
    \Delta S&=\frac{2\pi }{3R}\sum_{n,m_x,m_t=0}^\infty R^{4+2m_x+2m_t+2n}\frac{1}{(2m_x)!}\frac{1}{(2m_t)!}\partial_{x_0}^{2m_x}\partial_{t_{E0}}^{2m_t}\Big(T^{(n)}_{t_Et_E}C^{tt}_{n,m_x,m_t}+T_{xx}^{(n)}C^{xx}_{n,m_x,m_t}\Big),
\end{align}
where
\begin{align}
    C^{tt}_{n,m_x,m_t}&=I_{m_x,m_t,n}-I_{m_x,m_t+1,n}-\frac{2m_x}{2m_t+1}I_{m_x,m_t+1,n}-\frac{2n(2n+3)}{(2m_t+1)(2m_x+1)}I_{m_x+1,m_t+1,n-1},\\
    C^{xx}_{n,m_x,m_t}&=I_{m_x,m_t,n}-I_{m_x+1,m_t,n}-\frac{2m_t}{2m_x+1}I_{m_x+1,m_t,n}-\frac{2n(2n+3)}{(2m_t+1)(2m_x+1)}I_{m_x+1,m_t+1,n-1}.
\end{align}
When $n\geq1$, we have $C^{tt}_{n,m_x,m_t}=C^{xx}_{n,m_x,m_t}=0$. But for $n=0$
\begin{align}
    C^{tt}_{0,m_x,m_t}&=\frac{3}{2}I_{m_x,m_t,1},\\
    C^{xx}_{0,m_x,m_t}&=\frac{3}{2}I_{m_x,m_t,1}.
\end{align}
Therefore,
\begin{align}
    \Delta S&=\frac{\pi}{R}\sum_{m_x,m_t=0}^{\infty}R^{4+2m_x+2m_t}\frac{1}{(2m_x)!}\frac{1}{(2m_t)!}\partial_{x_0}^{2m_x}\partial_{t_{E0}}^{2m_t}I_{m_x,m_t,1}\big(T^{(0)}_{t_E t_E}(t_{E0},x_0,y)+T^{(0)}_{xx}(t_{E0},x_0,y)\big)\notag\\
    &=\frac{\pi}{R}\sum_{m_x,m_t=0}^{\infty}\frac{1}{(2m_x)!}\frac{1}{(2m_t)!}\partial_{x_0}^{2m_x}\partial_{t_{E0}}^{2m_t}T^{(0)}_{yy}\int dx dt_E(R^2-t_E^2-x^2)x^{2m_x}t_E^{2m_t} \notag\\
    &=\Delta \langle H \rangle.
\end{align}
Thus, we derive $\Delta S=\Delta \langle H \rangle$ from the solutions of linearized Einstein's equations.

\subsubsection{Deriving linearized Einstein's equations from the entanglement first law in AdS$_4/$CFT$_3$}
To prove this, we assume that $\tilde{h}_{\mu\nu}$ (for $\mu,\nu=0,1,2$), which may not satisfy Einstein's equations, is a solution of
\begin{align}
\Delta S(\tilde{h}_{\mu\nu})=\Delta \langle H_1 \rangle,
\end{align}
and we have another $h_{\mu\nu}$ to be the solution of eq.(\ref{eq for h}) that satisfies
\begin{align}
\Delta S(h_{\mu\nu})=\Delta \langle H_2 \rangle.
\end{align}
We define the difference between the two metric perturbations as $\Delta_{\mu\nu} = \tilde{h}_{\mu\nu} - h_{\mu\nu}$. 
 Since both expressions must hold on the boundary, which means that $\Delta \langle H_1 \rangle = \Delta \langle H_2 \rangle$, then we have
 \begin{align}
     \Delta_{\mu\nu}(z=0,t,x,y)=0 \label{delta at boundary},
 \end{align}
 which is the boundary conditions for $\Delta_{\mu\nu}$. And away from the boundary, the difference must satisfy
 \begin{align}
     0=\Delta S(\Delta_{\mu\nu}) \label{Deltamunu equation}.
 \end{align}
We assume that the subsystem is centered at $(t_0, x_0)$. Our goal is to show that this difference vanishes for arbitrary values of $t_0$, $x_0$, and $T_0$ in all Lorentz frames. Thus, the metric we consider in this part has the standard Lorentzian signature $\{-1,1,\dots,1\}$. We first consider this result in a fixed frame. 
Eq.(\ref{Deltamunu equation}) can be explicitly written as 
\begin{align}
    0=
    &\int_{(t-t_0)^2-(x-x_0)^2\geq T_0^2} dt dx\big(-(t^2-T_0^2)\Delta_{tt}(z,t-t_0,x-x_0,y)+(-T_0^2-x^2)\Delta_{xx}(z,t-t_0,x-x_0,y),\notag\\
    =&\sum_{n,m_x,m_t=0}^{\infty}\int dtdx(t^2-x^2-T_0^2)^\frac{n}{2}\bigg(\frac{1}{(2m_x)!}\frac{1}{(2m_t)!}\partial_{t_0}^{2m_x}\partial_{x_0}^{2m_x}\Delta_{tt}^{(n)}(t_0,x_0,y)x^{2m_x}t^{2m_t}(-t^2+T_0^2)\notag\\
    &-\frac{1}{(2m_x)!}\frac{1}{(2m_t)!}\partial_{t_0}^{2m_x}\partial_{x_0}^{2m_x}\Delta_{xx}^{(n)}(t_0,x_0,y)x^{2m_x}t^{2m_t}(x^2+T_0^2)\notag\\
    &+2\frac{1}{(2m_x+1)!}\frac{1}{(2m_t+1)!}\partial_{t_0}^{2m_x+1}\partial_{x_0}^{2m_x+1}\Delta_{tx}^{(n)}(t_0,x_0,y)x^{2m_x+2}t^{2m_t+2}\bigg), \label{equ for deltamunu}
\end{align}
where we have used $\Delta_{\mu\nu}=\sum_{n=0}^{\infty}z^n \Delta_{\mu\nu}^{(n)}$ and expanded it at point $(t_0,x_0)$. The right side of this equation can be expanded in powers of $T_0$ using
\begin{align}
K_{n,m_x,m_t}&=\int_{t^2-x^2\geq T_0^2} dtdx (t^2-x^2-T_0^2)^{\frac{n}{2}}t^{2m_t}x^{2m_x},\notag\\
&=\frac{1}{2}i^{n}T_0^{2+2m_x+2m_t+n}L_{n,m_x,m_t}.
\end{align}
This result is computed in the Euclidean regime (after a Wick rotation), and the value of the original Lorentzian integral is defined as its analytic continuation back to real time. The constant factor is defined as $L_{n,m_x,m_t}=\frac{\Gamma(1+m_t+m_x)\Gamma(\frac{n}{2}+1)}{\Gamma(2+m_x+m_t+\frac{n}{2})}\int \cosh^{2m_t}{\theta }\sinh^{2m_x}{\theta }d\theta$. 

At order $T_0^{N+4}$, where $N=n+2m_x+2m_t$, the vanishing coefficients and the relation $L_{N,0,1}+L_{N,1,0}=0$ (which can be easily verified in Euclidean spacetime) allow us to rewrite eq.(\ref{equ for deltamunu}) as follows,
\begin{align}
    \Delta_{tt}^{(N)}-\Delta_{xx}^{(N)}=&\sum_{(m_t,m_x)\ne0}P_{tt}^{N,m_x,m_t}\partial_{t_0}^{2m_t}\partial_{x_0}^{2m_x}\Delta_{tt}^{(N-2m_x-2m_t)}-P_{xx}^{N,m_x,m_t}\partial_{t_0}^{2m_t}\partial_{x_0}^{2m_x}\Delta_{xx}^{(N-2m_x-2m_t)}\notag\\
    &+\big(P_{xt}^{N,m_x-1,m_t}\partial_{x_0}^{2m_x-1}\partial_{t_0}^{2m_t+1}+P_{xt}^{N,m_x,m_t-1}\partial_{x_0}^{2m_x+1}\partial_{t_0}^{2m_t-1}\big)\Delta_{tx}^{(N-2m_x-2m_t)} ,\label{at order N}
\end{align}
in which the coefficients are 
\begin{align}
    P^{N,m_x,m_t}_{tt}&=-i^{n-N}\frac{1}{(2m_x)!}\frac{1}{(2m_t)!}\frac{L_{n,m_x,m_t}-L_{n,m_x,m_t+1}}{L_{N,0,0}-L_{N,0,1}},\\
    P_{xx}^{N,m_x,m_t}&=-i^{n-N}\frac{1}{(2m_x)!}\frac{1}{(2m_t)!}\frac{L_{n,m_x,m_t}+L_{n,m_x+1,m_t}}{L_{N,0,0}-L_{N,0,1 }},\\
    P_{tx}^{N,m_x,m_t}&=-i^{n-N}\frac{1}{(2m_x+1)!}\frac{1}{(2m_t+1)!}\frac{L_{n,m_x+1,m_t+1}}{L_{N,0,0}-L_{N,0,1}}.
\end{align}
The first few equations of eq.(\ref{at order N}) give
\begin{align}
    \Delta_{tt}^{(0)}-\Delta_{xx}^{(0)}&=0,\notag\\
    \Delta_{tt}^{(1)}-\Delta_{xx}^{(1)}&=0,\notag\\
    \Delta_{tt}^{(2)}-\Delta_{xx}^{(2)}&=P_{tt}^{2,1,0}\partial_{x_0}^2\Delta_{tt}^{(0)}+P_{tt}^{2,0,1}\partial_{t_0}^2\Delta_{tt}^{(0)}-P_{xx}^{2,1,0}\partial_{x_0}^2\Delta_{xx}^{(0)}-P_{xx}^{2,0,1}\partial_{t_0}^2\Delta_{xx}^{(0)}+(P_{tx}^{2,1,0}+P_{tx}^{2,0,1})\partial_{t_0}\partial_{x_0}\Delta_{tx}^{(0)},\notag\\
    \Delta_{tt}^{(3)}-\Delta_{xx}^{(3)}&=P_{tt}^{3,1,0}\partial_{x_0}^2\Delta_{tt}^{(1)}+P_{tt}^{3,0,1}\partial_{t_0}^2\Delta_{tt}^{(1)}-P_{xx}^{3,1,0}\partial_{x_0}^2\Delta_{xx}^{(1)}-P_{xx}^{3,0,1}\partial_{t_0}^2\Delta_{xx}^{(1)}+(P_{tx}^{3,1,0}+P_{tx}^{3,0,1})\partial_{t_0}\partial_{x_0}\Delta_{tx}^{(1)}\label{example}.
\end{align}
These equations fully determine $\Delta_{tt}-\Delta_{xx}$ because higher-order terms can be expressed in terms of $\Delta^{(0)}$ and $\Delta^{(1)}$. Note that the constraint eq.(\ref{delta at boundary}) determines the elements $\Delta_{\mu\nu}^{(0)}$, but the remaining elements of $\Delta_{\mu\nu}^{(1)}$ are unconstrained. 

Requiring eq.(\ref{equ for deltamunu}) to hold in all Lorentz frames imposes further constraints, resulting in equations analogous to eq.(\ref{at order N}). Let us consider, specifically, a general frame of reference obtained via the boost
 \begin{align}
     \Lambda&=\begin{pmatrix}
         \gamma&\gamma\beta_x &\gamma\beta_y\\
         \gamma\beta_x &1+\beta _x^2\frac{\gamma^2}{\gamma+1}& \beta_x \beta_y\frac{\gamma^2}{\gamma+1}\\
         \gamma\beta_y & \beta_x\beta_y\frac{\gamma^2}{\gamma+1}&1+\beta_y^2\frac{\gamma^2}{\gamma+1}
     \end{pmatrix}.
 \end{align}
 Then the left side of eq.(\ref{at order N}) becomes
 \begin{align}
    &\Lambda_t^{\;\mu}\Lambda_t^{\;\nu}\Delta_{\mu\nu}-\Lambda_x^{\;\mu}\Lambda_x^{\;\nu}\Delta_{\mu\nu}\notag\\
    &=(1+\beta_y^2+\beta_y^2\beta_x^2)\Delta_{tt}+(\beta_y^2+\beta^2\beta_y^2-\frac{1}{4}\beta_x^2\beta_y^2)\Delta_{yy}+(-1+\frac{1}{4}\beta_y^2\beta_x^2)\Delta_{xx}\notag\\
    &+\beta_x\beta_y^2\Delta_{tx}+(2\beta_y+\beta_y^3+\beta^2\beta_y)\Delta_{ty}
    +(\beta_x\beta_y+\frac{5}{4}\beta_x\beta_y\beta^2\notag-\frac{1}{2}\beta_x^3\beta_y)\Delta_{xy},\\\label{after boost}
 \end{align}
 in which we have expanded $\gamma$ in terms of $\beta=\sqrt{\beta_x^2+\beta_y^2}$ and keep terms up to $O(\beta^4)$. The general version of the second equation in eq.(\ref{example}) becomes 
 \begin{align}
     \Delta_{tt}^{(1)}-\Delta_{xx}^{(1)}+2\beta_y\Delta_{ty}^{(1)}+\beta_x\beta_y\Delta_{xy}^{(1)}+\beta_y^2(\Delta_{yy}^{(1)}+\Delta_{tt}^{(1)})+\beta_y^2(\beta_x\Delta_{tx}^{(1)}+\beta_y\Delta_{ty}^{(1)})+\beta^2\beta_y\Delta_{ty}^{(1)}\notag\\
     +\frac{1}{4}\beta_y^2\beta_x^2(\Delta_{xx}^{(1)}-\Delta_{yy}^{(1)}+4\Delta_{tt}^{(1)})+(\frac{5}{4}\beta_x\beta_y\beta^2-\frac{1}{2}\beta_x^3\beta_y)\Delta_{xy}^{(1)}&=0.
 \end{align}
 At each order of $\beta$, we have
 \begin{align}
     \Delta_{tt}^{(1)}-\Delta_{xx}^{(1)}&=0,\notag\\
     \Delta_{ty}^{(1)}&=0,\notag\\
     \Delta_{xy}^{(1)}=0\Delta_{yy}^{(1)}+\Delta_{tt}^{(1)}&=0,\notag\\
      \Delta_{tx}^{(1)}&=0,\notag\\
      \Delta_{xx}^{(1)}-\Delta_{yy}^{(1)}+4\Delta_{tt}^{(1)}&=0 .\label{delta 1=0}
 \end{align}
Therefore, we obtain $\Delta_{\mu\nu}^{(1)}=0$. For higher-order terms, we proceed by induction. Suppose that $\Delta_{\mu\nu}^{(n-1)}=0$; then at $n$-th order, eq.(\ref{after boost}) becomes
\begin{align}
\Delta_{tt}^{(n)}-\Delta_{xx}^{(n)}+2\beta_y\Delta_{ty}^{(n)}+\beta_x\beta_y\Delta_{xy}^{(n)}+\beta_y^2(\Delta_{yy}^{(n)}+\Delta_{tt}^{(n)})+\beta_y^2(\beta_x\Delta_{tx}^{(n)}+\beta_y\Delta_{ty}^{(n)})+\beta^2\beta_y\Delta_{ty}^{(n)}\notag\\
     +\frac{1}{4}\beta_y^2\beta_x^2(\Delta_{xx}^{(n)}-\Delta_{yy}^{(n)})+\beta^4\Delta_{tt}^{(n)}+(\frac{5}{4}\beta_x\beta_y\beta^2-\frac{1}{2}\beta_x^3\beta_y)\Delta_{xy}^{(n)}=f(\Delta_{\mu\nu}^{(n-1)})=0.
\end{align}
Repeating the same analysis as above, we find $\Delta_{\mu\nu}^{(n)}=0$, which means that the difference vanishes at every order. Thus we can conclude $\Delta_{\mu\nu}=0$. Consequently, we have completed the proof that the variation of the metric $h_{\mu\nu}$ which satisfies the entanglement first law also satisfies the linearized Einstein equations.

\subsection{The cases of arbitrary $d>3$}
For an arbitrary number of spacetime dimensions, Einstein's equations of metric in eq.(\ref{metric for FG expansion}) are \cite{Blanco:2013joa}
\begin{align}
T_\mu^{(n)\mu}=0,\quad \partial_{\mu}T^{(n)\mu\nu}=0,\quad
T_{\mu\nu}^{(n)}=\frac{(-1)^n\Gamma(d/2+1)}{2^{2n}n!\Gamma(d/2+n+1)}\Box^nT^{(0)}_{\mu\nu}.
\end{align}
In general, it is more convenient to discuss these equations in momentum space. Without loss of generality, we set the spatial momentum along the $x_{d-1}$ direction such that $T^{(0)}_{\mu\nu}=\hat{T}_{\mu\nu}e^{ip_{0}t_E}e^{-ip_{d-1}x_{d-1}}$. From the conservation and tracelessness of $T^{(0)}_{\mu\nu}$, we obtain the following equations
\begin{align}
\hat{T}_{(d-1)(d-1)} &= \pqty{\frac{p_0}{p_{d-1}}}^2 \hat{T}_{00},\\
\hat{T}^{i}_{i} &= \hat{T}_{(d-1)(d-1)}-\hat{T}_{00},\quad i=1,\dots,d-2,\\
T_{\mu\nu}^{(n)}&=e^{ip_0t_E-ip_{d-1}x_{d-1}}p^{2n}\frac{\Gamma(d/2+1)}{2^{2n}n!\Gamma(d/2+n+1)}\hat{T}_{\mu\nu},\quad \mu,\nu=0,1,\dots,d-2,
\end{align}
where $p^2=p_0^2-p_{d-1}^2$.

\subsubsection{$\Delta S=\Delta \langle H\rangle$ from linearized Einstein's equations}
Although the method in this part is analogous to the lower-dimensional case, it differs slightly because Einstein's equations become more complicated. Specifically, expanding the stress tensor around a center point $(t_{E0}, x_{01}, \dots, x_{0(d-1)})$ would lead to significant complexity when replacing terms such as $\partial_{t_E}\partial_{x_1}\dots \partial_{x_{d-2}} T_{0i}$. Thus, we will adopt an alternative approach to achieve our goal.

Let us consider the modular Hamiltonian in the original AdS$_{d+1}$ spacetime with signature $\{-1,1,\dots,1\}$ first, which is
\begin{align}
\Delta \langle H\rangle&=-\pi \int_{t^2-\sum_{i=1}^{d-2}x_i^2\geq T_0}d^{d-1}x \frac{T_0^2-\rho^2}{T_0} T^{(0)L}_{(d-1)(d-1)}
\notag\\
&=\frac{2\pi}{T_0 d}\hat{T}^L_{(d-1)(d-1)} e^{-i p_{d-1}x_{d-1}}\Omega_{d-2}\frac{1}{(d-2)\sqrt{\pi}}\left(\frac{-ip_0+\epsilon}{2T_0}\right)^{\frac{-1-d}{2}}K_{\frac{d+1}{2}}[T_0(-ip_0+\epsilon)]\Gamma\left(1+\frac{d}{2}\right), \label{mh in d}
\end{align}
where the superscript $L$ denotes the original Lorentz spacetime. And to regulate the integral, we insert a factor of 
$e^{-\epsilon t}$. Taking the limit $\epsilon\to 0$ then yields a divergent result, as expected. However, we will see that although both the modular Hamiltonian and the variation of entanglement entropy are divergent, they exhibit the same type of divergence. Specifically, both divergence are caused by the infinite volume (which would be an IR divergence).

The calculation of the deviation of entanglement entropy is performed in the spacetime obtained after the double Wick rotation, at the end, we will apply the inverse transformation to compare the result with eq.(\ref{mh in d}). The expression for $\Delta S$ is
\begin{align}
\Delta S&= \frac{ 2\pi}{d R}\int d^{d-1}x\sum_{n=0}z^{2n}\pqty{(R^2-t_E^2)T^{(n)}_{00}+(-x^i x^j+R^2g^{ij})T^{(n)}_{ij}-2x^i t_ET^{(n)}_{i0}},\label{deltaSe}
\end{align}
where $z^2+r^2=R^2$ and $r^2=t_E^2+\sum_{i=1}^{d-2}x_i^2$. This integral is symmetric under rotations that leave $x_{d-1}$ fixed. Subsequently, terms with $\mu \neq \nu$ in $\hat{T}_{\mu\nu}x^\mu x^\nu$ vanish upon integration, and for $i=j=1,\dots,d-2$, all integrals are equal. We can therefore replace
\begin{align}
&\sum_{i=0}^{d-2}(-x^i x^i+R^2g^{ii})\hat{T}_{ii},\notag\\
& \rightarrow \hat{T}_{(d-1)(d-1)}\bigg(R^2-t_E^2+\big(\frac{p}{p_0}\big)^2t_E^2+\frac{t_E^2-r^2}{d-2}\big(\frac{p}{p_0}\big)^2 \bigg).
\end{align}

This implies that
\begin{align}
\Delta S&=\frac{2\pi}{R d}\hat{T}_{(d-1)(d-1)} e^{-i p_{d-1}x_{d-1}}\int d^{d-1}x e^{ip_0t_E}\sum_{n=0}^\infty\frac{\Gamma(d/2+1)}{2^{2n}n!\Gamma(d/2+n+1)}(|p|z)^{2n}\notag\\
&  \bigg((R^2-t_E^2)+\big(\frac{p}{p_0}\big)^2t_E^2+\frac{t_E^2-r^2}{d-2}\big(\frac{p}{p_0}\big)^2 \bigg).\label{after replace}
\end{align}
Since the expression for the modular Hamiltonian is independent of $p^2$, the coefficients of $p^{2k}$ for $k \neq 0$ in eq.(\ref{after replace}) must vanish. This fact can be proved by taking the integration; the coefficient of $ p^n$ for $ n\neq 0$ vanishes (see more details in appendix \ref{prove the coefficient}). Therefore, we can take $p\to 0$ and only consider the $n=0$ term, namely,
\begin{align}
\Delta S&=\frac{2\pi}{R d}\hat{T}_{(d-1)(d-1)} e^{-i p_{d-1}x_{d-1}}\int d^{d-2}x dt_Ee^{ip_0t_E}(R^2-t_E^2). \label{p^0}
\end{align}

Now applying the inverse double Wick rotation $t_E=it, R=iT_0, r^2=-\rho^2, p_0\to -ip_0, x_{d-1}\to-ix_{d-1}, p_{d-1}\to ip_{d-1}$ , we obtain
\begin{align}
\Delta S&=\frac{2\pi}{T_0 d}\hat{T}^L_{(d-1)(d-1)} e^{-i p_{d-1}x_{d-1}}\int d^{d-2}x dt e^{ip_0t}(-T_0^2+t^2)\notag\\
&=\frac{2\pi}{T_0 d}\hat{T}^L_{(d-1)(d-1)} e^{-i p_{d-1}x_{d-1}}\Omega_{d-2}\frac{1}{(d-2)\sqrt{\pi}}\bigg(\frac{-ip_0+\epsilon}{2T_0}\bigg)^{\frac{-1-d}{2}}K_{\frac{d+1}{2}}[T_0(-ip_0+\epsilon)]\Gamma\left(1+\frac{d}{2}\right)\notag\\
&=\Delta\langle H\rangle,
\end{align}
in which we have utilized a similar trick to regulate the integral. Note that taking the same $\epsilon\to 0$ limit again yields a divergent result caused by the infinite volume, which is of the same type as the divergence in the modular Hamiltonian.

\subsubsection{Linearized Einstein's equations from $\Delta S=\Delta \langle H\rangle$}
We again assume that there exist two solutions satisfying the entanglement first law, such that one of them satisfies Einstein's equations while the other does not. We define their difference as $\Delta_{\mu\nu}$, which must satisfy eq.\eqref{delta at boundary} and eq.\eqref{Deltamunu equation}. The procedure of the proof follows the same logic as in lower dimensions. Suppose that the subsystem is centered at $(t_0,x_{01},\dots,x_{0(d-2)})$ and has a radius $T_0$. We need to show that $\Delta_{\mu\nu}=0$ is independent of the choices of the center coordinates $(t_0,\dots,x_{0(d-2)})$, the radius $T_0$, and the reference frame.

In a fixed frame, we have
\begin{align}
0=
&\sum_{n=0}^\infty\int d^{d-1}x z^n\Big((T_0^2-t^2)\Delta^{(n)}_{tt}(x_\mu-x_{0\mu},x_{d-1})\notag\\&-\sum_{i=1}^{d-2}(T_0^2+x_i^2)\Delta^{(n)}_{ii}(x_\mu-x_{0\mu},x_{d-1})+2\sum_{i=1}^{d-2}tx_i\Delta^{(n)}_{tx_i}(x_\mu-x_{0\mu},x_{d-1})\Big),\notag\\
=&\sum_{n,\{m_\mu=0\}}^\infty\int d^{d-1}xz^n\Bigg(\prod_{\mu=0}^{d-2}(x_{\mu})^{2m_\mu}\partial_{0\mu}^{2m_\mu}\frac{1}{(2m_\mu)!}\notag\\
&\times\bigg(\Delta_{tt}^{(n)}(x_{0\mu},x_{d-1})(T_0^2-t^2)-\sum_{i=1}^{d-2}\Delta_{ii}^{(n)}(x_{0\mu},x_{d-1})(T_0^2+x_i^2)\bigg)\notag\\
&+2\sum_{i=1}^{d-2}\prod_{\mu\neq 0,i}^{d-2}\frac{1}{(2m_\mu)!}\frac{1}{(2m_i+1)!}\frac{1}{(2m_t+1)!}\partial_{0\mu}^{2m_\mu}\partial_{0t}^{2m_t+1}\partial_{0i}^{2m_i+1}\notag\\&\times (x_\mu)^{2m_\mu}t^{2m_t+2}x_i^{2m_i+2}\Delta_{ti}^{(n)}(x_{0\mu},x_{d-1})\Bigg),\label{fixed frame d}
\end{align}
where $\partial_{0\mu}\equiv\frac{\partial}{\partial x^\mu_0}$ and $\{m_\mu\}$ represents the index $\{m_t,m_1,\dots,m_{d-2}\}$. The integration can be replaced by
\begin{align}
K_{n,\{m_\mu\}}&=\int d^{d-1}x (\rho^2-T_0^2)^{\frac{n}{2}}\prod_{\mu=0}^{d-2}(x_\mu)^{2m_\mu},\notag\\
&=L_{n,\{m_\mu\}}i^n T_0^{-1+d+n+2\sum_{\mu=0}^{d-2}m_\mu},
\end{align}
(where $\rho^2=t^2-\sum_{i=1}^{d-2}x_i^2$ ) with the constant factor
\begin{align}
L_{n,\{m_\mu\}}&=\frac{\Gamma(-\frac{1}{2}+\frac{d}{2}+\sum_{\mu=0}^{d-2}m_\mu)\Gamma(1+\frac{n}{2})}{\Gamma\big(\frac{1+d+n}{2}+\sum_{\mu=0}^{d-2}m_\mu\big)}\Theta_{\{m_\mu\}},
\end{align}
 and
\begin{align}
\Theta_{\{m_{\mu}\}}&=\int (\cosh{\theta})^{2m_{t}}(\sinh{\theta}\cos{\phi_1})^{2m_1}\dots(\sinh{\theta}\sin{\phi_1}\dots\sin{\phi_{d-3}} )^{2m_{d-2}} d\theta d\Omega_{d-3}.
\end{align}
in which $d\Omega_{d-3}=(\sin{\phi_1})^{d-4}(\sin{\phi_2})^{d-5}\dots \sin{\phi_{d-4}}d\phi_1d\phi_2\dots d\phi_{d-3}$, $\phi_{i\neq d-3}\in[0,\pi)$, $\phi_{d-3}\in[0,2\pi)$.
Eq.(\ref{fixed frame d}) can be rewritten more simply as
\begin{align}
0=&\frac{1}{2}\sum_{n,\{m_\mu=0\}}i^nT_0^{d+1+n+2\sum_{\mu=0}^{d-2}m_\mu}\prod_{\mu=0}^{d-2}\frac{1}{(2m_\mu)!}\partial_{0\mu}^{2m_\mu}\bigg(\Delta_{tt}^{(n)}\pqty{L_{n,\{m_\mu\}}-L_{n,\{m_t+1,m_{\mu\neq 0}\}}}\notag\\&-\sum_{i=1}^{d-2}\Delta_{ii}^{(n)}\pqty{L_{n,\{m_\mu\}}+ L_{n,\{m_{\mu\neq i},m_i+1\}}}\notag\\
&+\sum_{i=1}^{d-2}T_0^2L_{n,\{m_{\mu\neq 0,i},m_t+1,m_i+1\}}\frac{1}{2m_t+1}\frac{1}{2m_i+1}\partial_{0t}\partial_{0i}\Delta_{ti}^{(n)}\bigg).
\end{align}
Considering the order up to $T^{N+d+1}$, $N=n+\sum_{\mu=0}^{d-2}m_\mu$, we have
\begin{align}
&\Delta_{tt}^{(N)}-\sum_{i=1}^{d-2}C_i\Delta_{ii}^{(N)}\notag\\
&=\sum_{N=0,\{m_\mu\neq 0\}}\prod_{\mu=0}^{d-2}\partial_{0\mu}^{2m_\mu}\pqty{P^{N,\{m_\mu\}}_{tt}\Delta_{tt}^{(N-2\sum_{\mu=0}^{d-2}m_\mu)}-\sum_{i=1}^{d-2}P_{ii}^{N,\{m_\mu\}}\Delta_{ii}^{(N-\sum_{\mu=0}^{d-2}m_\mu)}}\notag\\
&+\sum_{i=1}^{d-2}\prod_{\mu\neq0,i}^{d-2}\partial_{0\mu}^{2m_\mu}\pqty{P^{N,\{m_t-1,m_{\mu\neq0}\}}_{ti}\partial_{0i}^{2m_i+1}\partial_{0t}^{2m_t-1}+P^{N,\{m_{\mu\neq i},m_i-1\}}_{ti}\partial_{0i}^{2m_i-1}\partial_{0t}^{2m_t+1}}\Delta_{it}^{(N-\sum_{\mu=0}^{d-2}m_\mu)},\label{equ for delta after boost at d}
\end{align}
where the coefficients are as follows.
\begin{align}
C_i&=\frac{L_{n,\{m_{\mu}=0\}}+L_{n,\{m_{\mu\neq i}=0,m_i=1\}}}{L_{n,\{m_{\mu}=0\}}-L_{n,\{m_t=1,m_{\mu\neq 0}=0\}}},\\
P_{tt}^{N,\{m_\mu\}}&=-i^{n-N}\prod_{\mu=0}^{d-2}\frac{1}{(2m_\mu)!}\frac{L_{n,\{m_\mu\}}-L_{n,\{m_t+1,m_{\mu\neq0}\}}}{L_{n,\{m_{\mu}=0\}}-L_{n,\{m_t=1,m_{\mu\neq 0}=0\}}},\\
P_{ii}^{N,\{m_\mu\}}&=-i^{n-N}\prod_{\mu=0}^{d-2}\frac{1}{(2m_\mu)!}\frac{L_{n,\{m_\mu\}}+L_{n,\{m_{\mu\neq i},m_i+1\}}}{L_{n,\{m_{\mu}=0\}}-L_{n,\{m_t=1,m_{\mu\neq 0}=0\}}},\\
P_{it}^{N,\{m_\mu\}}&=-\frac{2}{(d-2)}i^{n-N}\prod_{\mu\neq 0,i}^{d-2}\frac{1}{(2m_\mu)!}\frac{1}{(2m_t+1)!}\frac{1}{(2m_i+1)!}\frac{L_{n,\{m_{\mu\neq 0,i},m_t+1,m_i+1\}}}{L_{n,\{m_{\mu}=0\}}-L_{n,\{m_t=1,m_{\mu\neq 0}=0\}}},
\end{align}
where the factor $\frac{2}{(d-2)}$ comes from: the order of $\Delta_{ti}$ is two orders higher than the other components, and there are $d-2$ different ways to redefine $m_{\mu}$ such that all components are of the same order. 
Moreover, the integral
$\Theta_{\{m_{\mu \neq i}=0,\,m_i=1\}}$ is invariant under rotations. To show this, let 
$\vec{n}=(n_1,\dots,n_{d-2})$ denote the unit direction vector on the sphere $S^{d-3}$, and let $R$ be a rotation matrix acting on 
$\vec{n}$. Then
\begin{align}
    \int_{S^{d-3}} n_i^{2}\,d\Omega_{d-3}
= \int_{S^{d-3}} (R\vec{n})_i^{\,2}\,d\Omega_{d-3}
= \int_{S^{d-3}} n_j^{2}\,d\Omega_{d-3}.
\end{align}
Therefore, we conclude that
\begin{align}
    C_i = C_j = C, \qquad \text{for all } i,j.
\end{align}

We then can write the first few equations of eq.(\ref{equ for delta after boost at d}) as
\begin{align}
    &\Delta_{tt}^{(0)}-\sum_{i=1}^{d-2}C_i\Delta_{ii}^{(0)}=0,\notag\\
    &\Delta_{tt}^{(1)}-\sum_{i=1}^{d-2}C_i\Delta_{ii}^{(1)}=0,\notag\\
    &\Delta_{tt}^{(2)}-\sum_{i=1}^{d-2}C_i\Delta_{ii}^{(2)}\notag\\=&P_{tt}^{2,\{m_t=1,m_{\mu\neq 0}=0\}}\partial_{0t}^2\Delta_{tt}^{(0)}+\sum_{i=1}^{d-2}P_{tt}^{2,\{m_{\mu\neq i}=0,m_i=1\}}\partial_{0i}^2\Delta_{tt}^{(0)}-\sum_{i=1}^{d-2}P_{ii}^{2,\{m_t=1,m_{\mu\neq 0}=0\}}\partial_{0t}^2\Delta_{ii}^{(0)}\notag\\
    &-\sum_{i=1}^{d-2}P_{ii}^{2,\{m_{\mu\neq i}=0,m_i=1\}}\partial_{0i}^2\Delta_{ii}^{(0)}+\sum_{i=1}^{d-2}\pqty{P_{ti}^{2,\{m_t=1,m_{\mu\neq 0}=0\}}\partial_{0t}^2+P^{2,\{m_{\mu\neq i}=0,m_i=1\}}_{ti}\partial_{0i}^2}\Delta_{ti}^{(0)},\notag\\
    &\Delta_{tt}^{(3)}-\sum_{i=1}^{d-2}C_i\Delta_{ii}^{(3)}\notag\\
    =&P_{tt}^{2,\{m_t=1,m_{\mu\neq 0}=0\}}\partial_{0t}^2\Delta_{tt}^{(1)}+\sum_{i=1}^{d-2}P_{tt}^{2,\{m_{\mu\neq i}=0,m_i=1\}}\partial_{0i}^2\Delta_{tt}^{(1)}-\sum_{i=1}^{d-2}P_{ii}^{2,\{m_t=1,m_{\mu\neq 0}=0\}}\partial_{0t}^2\Delta_{ii}^{(1)}\notag\\
    &-\sum_{i=1}^{d-2}P_{ii}^{2,\{m_{\mu\neq i}=0,m_i=1\}}\partial_{0i}^2\Delta_{ii}^{(1)}+\sum_{i=1}^{d-2}\pqty{P_{ti}^{2,\{m_t=1,m_{\mu\neq 0}=0\}}\partial_{0t}^2+P^{2,\{m_{\mu\neq i}=0,m_i=1\}}_{ti}\partial_{0i}^2}\Delta_{ti}^{(1)}. \label{explicit delta 1}
\end{align}
 \(\Delta_{\mu\nu}^{(0)}\) is fully fixed by the boundary condition. To obtain higher-order \(\Delta_{\mu\nu}^{(n)}\), considering a general boost in $d$-dimensional spacetime
\begin{align}
    \Lambda&=\begin{pmatrix}
        \gamma&\gamma\beta_1&\gamma\beta_2&\dots&\gamma\beta_{d-1}\\
        \gamma\beta_1 & 1+(\gamma-1)\frac{\beta_1^2}{\beta^2}&(\gamma-1)\frac{\beta_1\beta_2}{\beta^2}&\dots&(\gamma-1)\frac{\beta_1\beta_{d-1}}{\beta^2}\\
        \vdots&\vdots&\vdots&\ddots&\vdots\\
        \gamma\beta_{d-1}&(\gamma-1)\frac{\beta_{d-1}\beta_1}{\beta^2}&(\gamma-1)\frac{\beta_{d-1}\beta_2}{\beta^2}&\dots&1+(\gamma-1)\frac{\beta_{d-1}^2}{\beta^2}
    \end{pmatrix},
\end{align}
where $\beta^2=\sum_{j=1}^{d-1}\beta_i^2$. The left side of eq.(\ref{equ for delta after boost at d}) becomes
\begin{align}
\Lambda_t^{\;\mu}\Lambda_t^{\;\nu}\Delta_{\mu\nu}-\sum_{i=1}^{d-2}C_i\Lambda_i^{\;\mu}\Lambda_i^{\;\nu}\Delta_{\mu\nu}=M_{tt}\Delta_{tt}+2\sum_{j=1}^{d-1}M_{tj}\Delta_{tj}+\sum_{j,k=1}^{d-1}M_{jk}\Delta_{jk}.
\end{align}
Expanding $\gamma$ in $\beta$ and keeping terms up to $O(\beta^3)$ and after this boost, the second equation of eq.(\ref{explicit delta 1}) becomes
\begin{align}
    &\Delta_{tt}^{(1)}-C\sum_{i=1}^{d-2}\Delta_{ii}^{(1)}+2\beta_{d-1}\Delta_{t(d-1)}^{(1)}+(2-2C)\sum_{i=1}^{d-2}\beta_i\Delta_{ti}^{(1)}
    +\beta_{d-1}^2\pqty{\Delta_{tt}^{(1)}+\Delta_{(d-1)(d-1)}^{(1)}}\notag\\
    &+(2-C)\sum_{i=1}^{d-2}\beta_i\beta_{d-1}\Delta_{i(d-1)}^{(1)}+(1-C)\sum_{i=1}^{d-2}\beta_i^2\pqty{\Delta_{ii}^{(1)}+\Delta_{tt}^{(1)}}+(2-C)\sum_{i\neq j}^{d-2}\beta_i\beta_j \Delta_{ij}^{(1)}\notag\\
    &+\beta^2\pqty{(2-C)\beta_{d-1}\Delta_{t(d-1)}^{(1)}+(2-2C)\sum_{i=1}^{d-2}\beta_i\Delta_{ti}^{(1)}}+\beta_{d-1}^2\pqty{C\beta_{d-1}\Delta_{t(d-1)}^{(1)}+\sum_{i=1}^{d-2}\beta_i\Delta_{ti}^{(1)}}=0.
\end{align}
Since $(\beta_1,\dots,\beta_{d-1})$ are linearly independent and the coefficients of each power of $\beta$ must vanish, we can obtain the following equations 
\begin{align}
    \Delta_{tt}^{(1)}-C\sum_{i=1}^{d-2}\Delta_{ii}^{(1)}&=0,\notag\\
    \Delta_{tj}^{(1)}&=0,\quad {\rm for}\quad j=1,\dots,d-1.\notag\\
    \Delta_{tt}^{(1)}+\Delta_{(d-1)(d-1)}^{(1)}&=0,\notag\\
    \Delta_{ij}^{(1)}&=0,\quad {\rm for}\quad i\neq j, \ j=1,\dots,d-1,\ i=1,\dots,d-2.\notag\\
    \Delta_{ii}^{(1)}+\Delta_{tt}^{(1)}&=0.\quad {\rm for}\quad i=1,\dots,d-2.
\end{align}
When $C=1$ and $d=3$, we recover eq.(\ref{delta 1=0}). 
So far, we have shown that $\Delta_{\mu\nu}^{(1)}=0$. 
For higher-order terms, we can use induction to obtain $\Delta_{\mu\nu}^{(n)}=0$. 
Therefore, $\Delta_{\mu\nu}=0$, which completes our proof.

\section{Conclusions and discussions}\label{sec-conclu}
In this work, we investigated the entanglement first law for timelike entanglement entropy in relativistic quantum field theory and its holographic realization in asymptotically AdS$_{d+1}$ spacetime. For timelike-separated regions in the boundary CFT, we showed that a timelike entanglement first law can be formulated in terms of variations of the HTEE and the expectation value of the associated modular Hamiltonian. On the gravity side, we adopted the holographic prescription of~\cite{Heller:2024whi} in which HTEE is computed from codimension-two timelike extremal surfaces obtained by analytic continuation of the standard spacelike HEE setup. Imposing the timelike entanglement first law for all admissible timelike regions and using the area–entropy relation, we demonstrated that this first law is equivalent to the linearized Einstein's equations in the bulk asymptotically AdS spacetime, thereby strengthening the dynamical connections between entanglement and gravity.

For low-energy excitations of timelike subsystems, we calculated the variations of HTEE and energy for strip and hyperbolic regions and extracted the corresponding entanglement temperature. In the UV limit, the timelike entanglement temperature shows the expected universal behaviour, scaling inversely with the temporal size of the subsystem. At the same time, in the IR limit, it approaches the Hawking temperature up to a constant factor. We argued that attempts to remove this constant by redefining the entanglement first law or inserting ad hoc prefactors suffer from conceptual and technical issues. Motivated by this, we adopted a more robust formulation based on the non-negativity of relative entropy, yielding a general inequality between the modular Hamiltonian and the entanglement entropy that holds along the RG flow. For small perturbations, this inequality saturates and naturally reproduces a well-defined timelike entanglement first law, showing that the first law follows directly from the basic information-theoretic properties of quantum states.

To clarify the relation between the timelike entanglement first law and the bulk gravitational dynamics, we focused on hyperbolic timelike subregions. Using the modular Hamiltonian obtained via the double Wick rotation, we found that for thermal perturbations, the change in the modular Hamiltonian exactly matches the renormalized timelike entanglement entropy. Starting from the AdS$_4$/CFT$_3$ correspondence and then extending to arbitrary dimensions, we established the equivalence between the timelike entanglement first law and the linearized Einstein equations in asymptotically AdS$_{d+1}$ gravity. As in the spacelike case, this implies that perturbations of HTEE obey a differential equation equivalent to that of a scalar field on an emergent dS$_d$ spacetime, now interpretable as encoding a rotated time direction after double Wick rotation. This provided a timelike analogue of the emergence of spatial directions from entanglement and suggests a possible mechanism for the emergence of time in the gauge/gravity duality. Our construction also highlighted several features specific to timelike entanglement, including the central role of analytic continuation and double Wick rotation, as well as the interplay between causality, energy conditions, and the regularity of timelike bulk extremal surfaces.

There are several directions for future investigations. An obvious extension is to go beyond the linear regime and test whether higher-order variations of timelike entanglement entropy reproduce non-linear corrections to Einstein's equations. It would also be interesting to apply our framework to more general backgrounds, in particular time-dependent or cosmological ones, and to clarify the relation between timelike entanglement and dS or cosmological holography. Finally, incorporating bulk quantum corrections and quantum extremal timelike surfaces, as well as exploring modular flow and information-theoretic inequalities adapted to timelike regions, may yield further constraints on consistent bulk dynamics and help organize timelike entanglement into a more complete thermodynamics-like theory.

\section*{Acknowledgement}
We would like to thank Wu-Zhong Guo, Long Zhao, Zhixuan Zhao, and Hai-Qing Zhang for valuable discussions. J.R.S. was supported by the National Natural Science Foundation of China (No.~12475069) and Guangdong Basic and Applied Basic Research Foundation (No.~2025A1515011321). S. H. was supported by the National Natural Science Foundation of China (No. 12475053 and No. 12235016).

\appendix

\section{Modular Hamiltonian for hyperbolic subregion}\label{modularh}
\begin{figure}[h]
    \centering
    \includegraphics[width=0.5\linewidth]{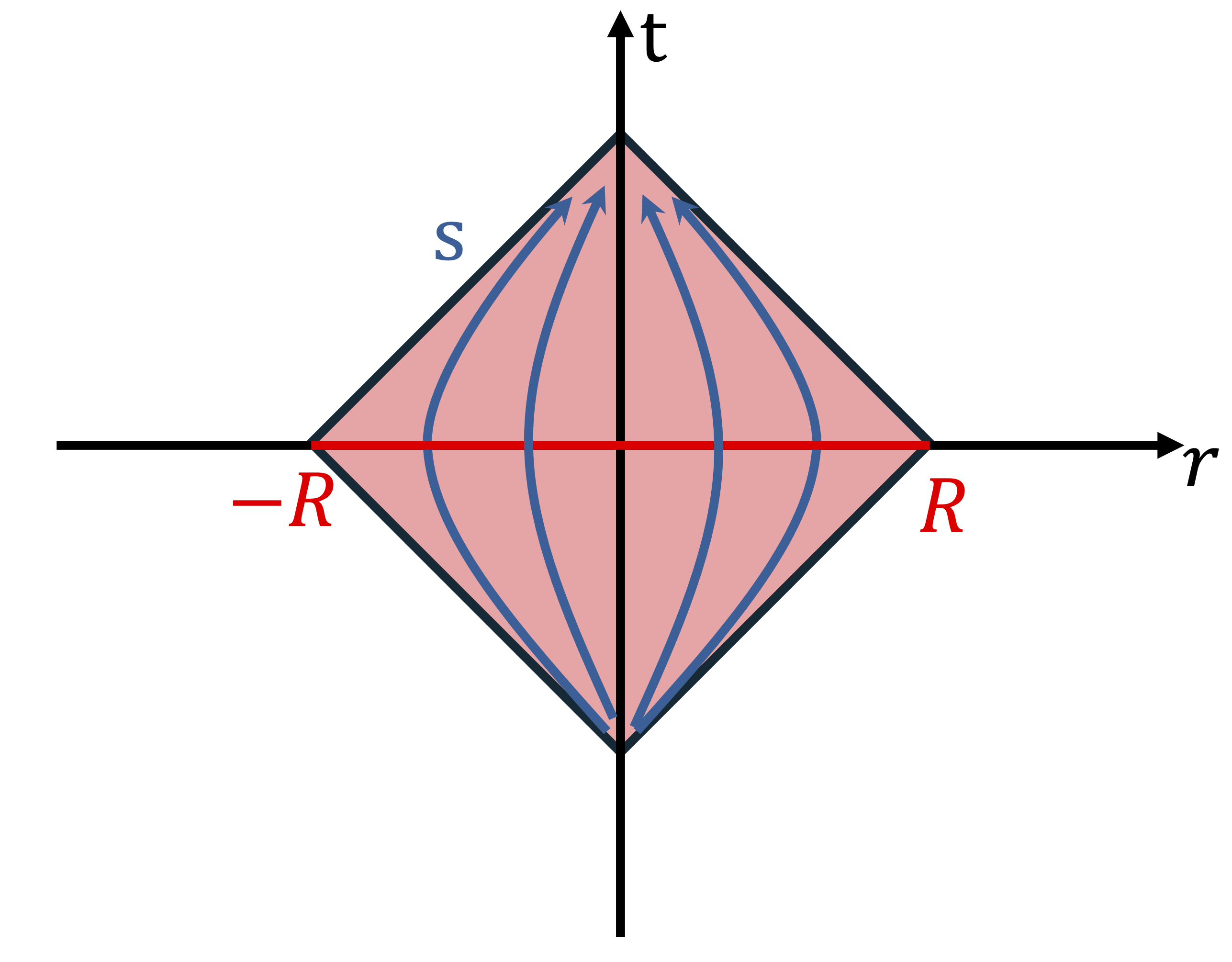}
    \caption{The domain of dependence of subregion $\mathcal{V}$ is the red region, and the modular flow is the blue line.}
    \label{DOD for V.png}
\end{figure}

Consider d-dimension CFT in Minkowski spacetime $\mathbb{R}^{1,d-1}$, and the subregion we are focused on is $\mathcal{V}:-R<r<R$ where $r=\sum_{i=1}^{d-1}x_i^2$. The domain of dependence $\mathcal{D}$ is shown in Fig.\ref{DOD for V.png}.
We can use the conformal transformation eq.(\ref{SCT}) which maps $\mathcal{D}$ into Rindler spacetime to get the modular Hamiltonian \cite{Casini:2011kv}
\begin{align}
      x^\mu&=\frac{X^\mu-(X\cdot X)C^\mu}{1-2(X\cdot C)+(X\cdot X)(C\cdot C)}+2R^2C^\mu, \label{SCT}\\
      H_{\mathcal{D}}&=2\pi \int_{\mathcal{V}}d^{d-1}x \frac{(R^2-r^2)}{2R}T^{00}(x),
\end{align}
here $C^\mu=(0,\frac{1}{2R},0,\dots,0)$. For a hyperbolic subregion, by using a double Wick rotation $t\to-it_E, x_{d-1}\to iy_{d-1}$, we derived the associated modular Hamiltonian.
\begin{figure}
    \centering
    \includegraphics[width=0.45\linewidth]{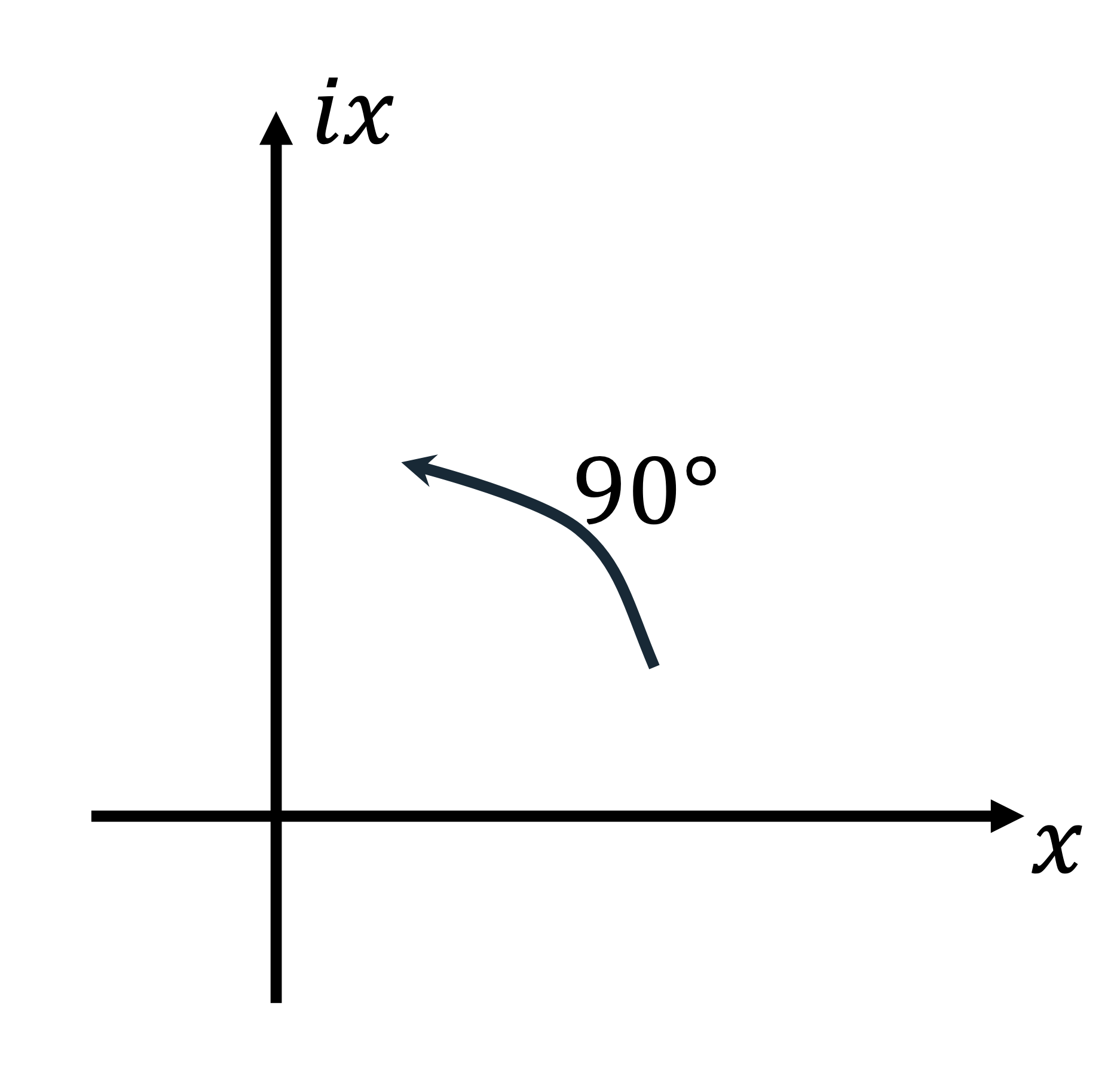}
    \caption{The geometric meaning of multiplying $i$.}
    \label{Wick rotation}
\end{figure}

\begin{figure}
    \centering
    \includegraphics[width=0.5\linewidth]{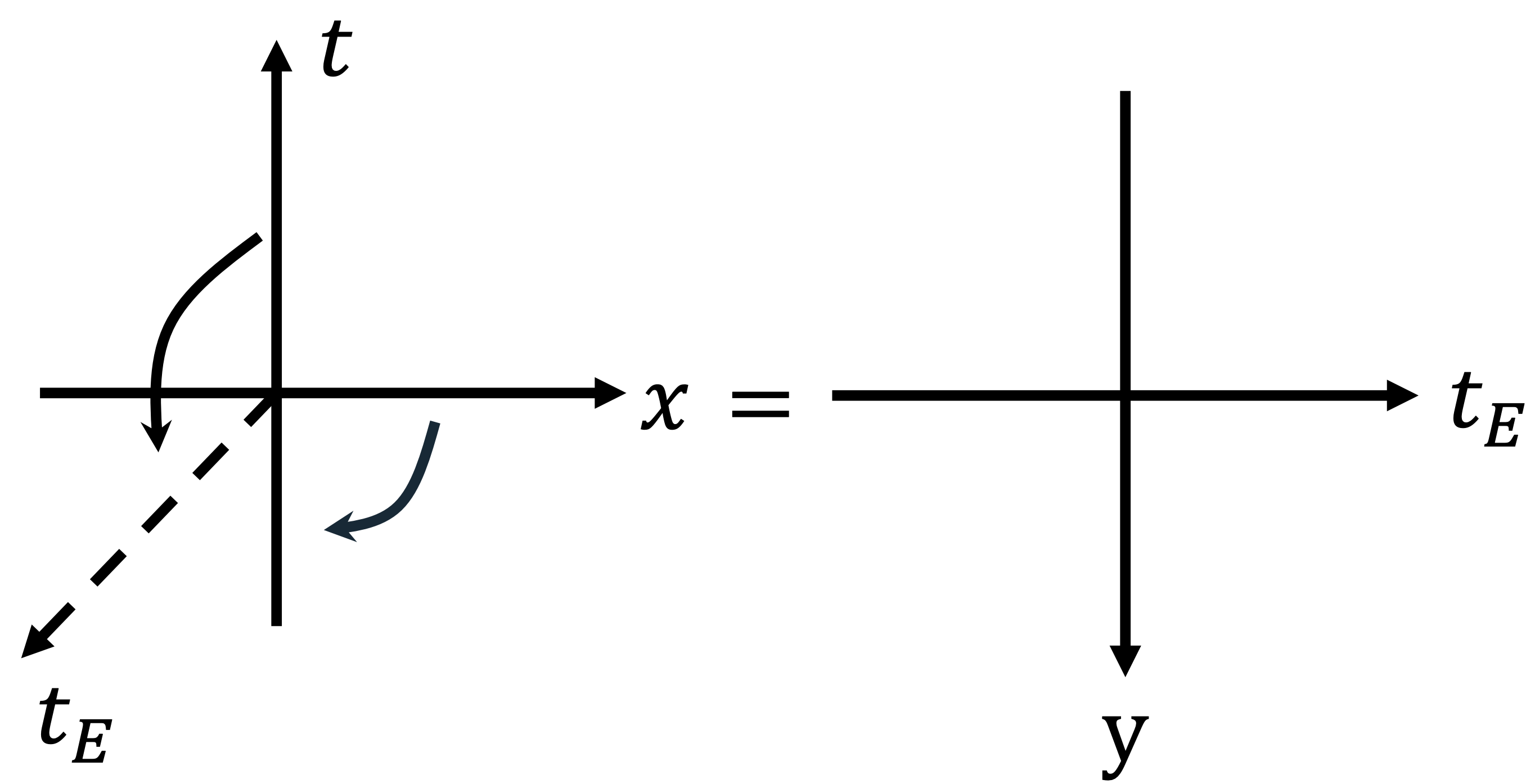}
    \caption{The diagram on the left illustrates the procedure of the double Wick rotation, while the one on the right shows the resulting coordinate system.}
    \label{double wick}
\end{figure}

To understand the geometric meaning of the double Wick rotation, we consider the case of $d=1+1$ dimensions. Since multiplication by $i$ corresponds to a $90^{\circ}$ anticlockwise rotation of the coordinates (as shown in Fig. \ref{Wick rotation}), performing a double Wick rotation is equivalent to rotating the time coordinate clockwise and the spatial coordinate ($x$) anticlockwise by $90^{\circ}$ each. After we do double Wick rotation, the final coordinate system is in Fig.\ref{double wick}.

We can conclude that the double Wick rotation does not transform the coordinates as $t \to x,\ x \to t$, but rather as $t \to x,\ x \to -t$.

\begin{figure}
    \centering
    \includegraphics[width=0.5\linewidth]{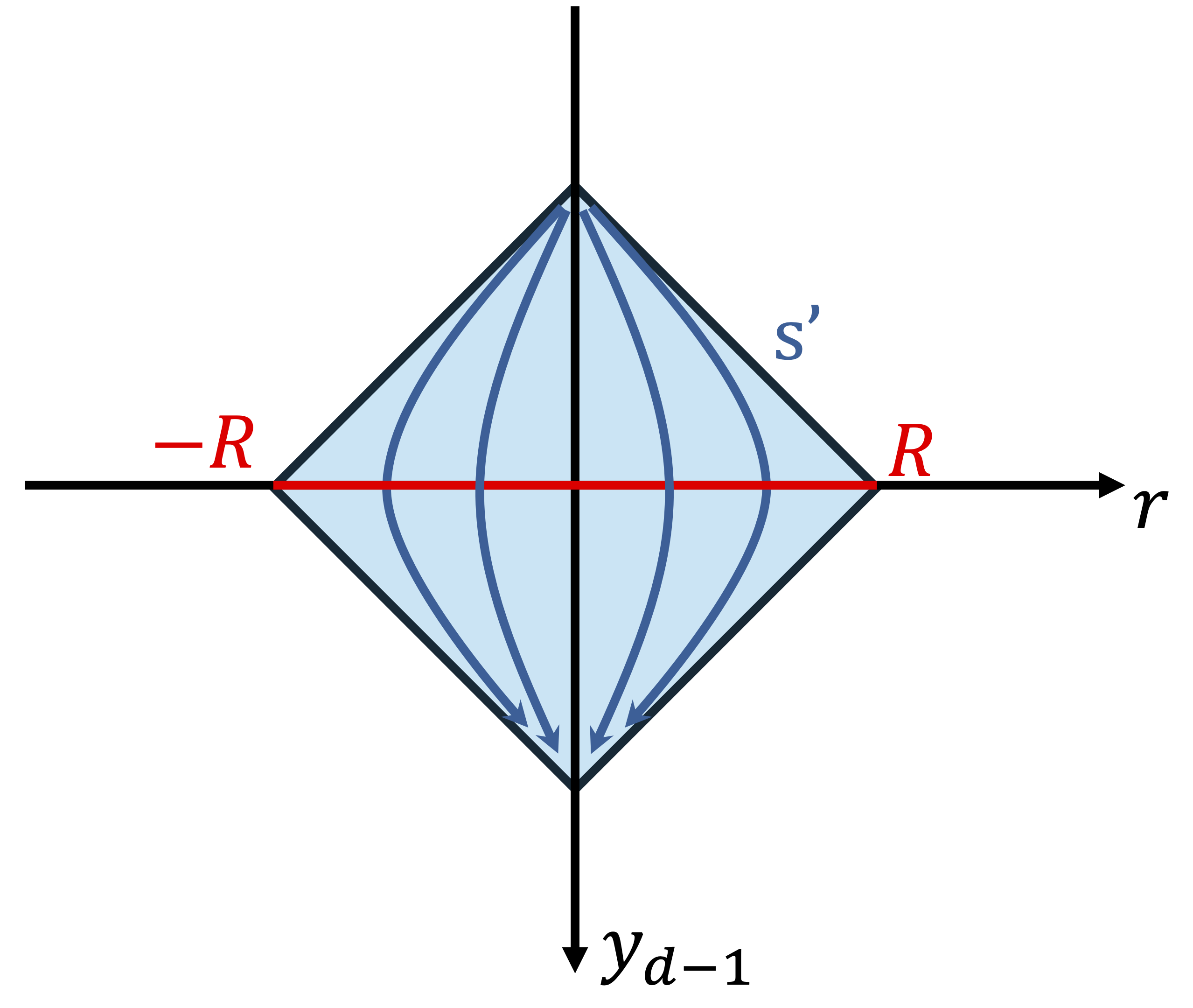}
    \caption{The blue region is the domain of dependence of $\mathcal{V'}_E$, and the modular flow is the blue curve. Here $r^2=t_E^2+\sum_{i=1}^{d-2}x_i^2$.}
    \label{after double wick rotation}
\end{figure}

In our case, the first Wick rotation $t \to -i t_E$ transforms the subregion $\mathcal{V'}$ into $\mathcal{V'}_E: t_E^2 + \sum_{i=1}^{d-2}x_i^2 \leq R^2$, where $R = iT_0$. Subsequently, applying the second Wick rotation $x_{d-1} \to i y_{d-1}$, the domain of dependence $\mathcal{D'}_E$ of $\mathcal{V'}_E$ is illustrated in Fig. \ref{after double wick rotation}. From this diagram, it is evident that the direction of the modular flow is the same as that of the ordinary ball-shaped subregion. Therefore, the modular Hamiltonian is
\begin{align}
    H_{\mathcal{D'}_E}&=i\partial_{s'},\notag\\
    &=i\partial_s,\notag\\
    &=2\pi \int_{t^2_E+\vec{x}^2\leq R^2}d^{d-2}x dt_E\frac{(R^2-r^2)}{2 R}T_E^{(d-1)(d-1)}.
\end{align}

Now take the inverse Wick rotation $t_E\to it, R\to iT_0,r^2\to -\rho^2$, where $\rho^2=t^2-\vec{x}^2$, we'll get the modular Hamiltonian for the domain of dependence $\mathcal{D'}$ of subregion $\mathcal{V'}$
\begin{align}
    H_{\mathcal{D'}}&=-2\pi \int_{t^2-\sum_{i=1}^{d-2}x_i^2\geq T_0^2} d^{d-2}x dt\frac{(T_0^2-\rho^2)}{2T_0}T^{(d-1)(d-1)} .
\end{align}
 The stress tensor satisfies \cite{Chen:2025leq} $T_E^{(d-1)(d-1)}=-T^{(d-1)(d-1)}$ (the minus sign is caused by the inverse Wick rotation related to $x_{d-1}$).
 
\section{Perturbative timelike entanglement entropy in Lorentzian spacetime}\label{real}
We consider the hyperbolic subregion defined by $t^2-\sum_{i=1}^{d-2}x_i^2\geq T_0^2$. The metric of perturbed AdS spacetime around the vacuum solution is given by
\begin{align}
    ds^2=\frac{1}{z^2}\left(dz^2+(\eta_{\mu\nu}+\delta g_{\mu\nu})dx^\mu dx^\nu\right).\quad \mu,\nu=0,1,\dots,d-1.
\end{align}
In Lorentzian spacetime, the extremal surface consists of three components \cite{Mollabashi:2021xsd,Li:2022tsv}: two spacelike surfaces $\rho^2-z^2=T_0^2$ which give the  real part and a timelike surface $z^2-\rho^2=C^2$  contributes to the imaginary part, where $C$ is an arbitrary positive constant. 
After perturbation, the contribution from the two spacelike surfaces are
\begin{align}
    \Delta S_{\rm real}&=\frac{4\pi}{T_0 d}\hat{T}_{(d-1)(d-1)} e^{-i p_{d-1}x_{d-1}}\int d^{d-1}x e^{ip_0t}\big(t^2-T_0^2\big)\notag\\
    &=\frac{4\pi}{T_0 d(d-2)}\hat{T}_{(d-1)(d-1)} e^{-i p_{d-1}x_{d-1}}\int_{T_0}^\infty dt e^{ip_0t}\pqty{t^2-T_0^2}^{d/2}.
\end{align}

While the imaginary contribution from the timelike surface is
\begin{align}
    \Delta S_{\rm imaginary}
    &=-i\frac{2 \pi }{d C}\hat{T}_{(d-1)(d-1)} e^{-i p_{d-1}x_{d-1}}\int d^{d-1}x e^{ip_0t}\big(t^2+C^2\big)\notag\\
    &=-i\frac{2 \pi }{d(d-2) C}\hat{T}_{(d-1)(d-1)} e^{-i p_{d-1}x_{d-1}}\int_{T_0}^\infty dt e^{ip_0t}\pqty{C^2+t^2}\pqty{t^2-C^2}^{-1+d/2}.
\end{align}
However, the modular Hamiltonian for this subregion is
\begin{align}
    \Delta\langle H\rangle&=-\pi \int_{t^2-\sum_{i=1}^{d-2}x_i^2\geq T_0^2}d^{d-1}x \frac{\big(T_0^2-\rho^2\big)}{T_0} T^{(0)}_{(d-1)(d-1)}\notag\\
     &=\frac{2\pi}{T_0 d(d-2)}\hat{T}_{(d-1)(d-1)} e^{-i p_{d-1}x_{d-1}}\Omega_{d-2}\int _{T_0}^{\infty} dt  e^{ip_0t}\pqty{t^2-T_0^2}^{d/2}.
\end{align}
Crucially, even we choose $C=T_0$, the change in the entanglement entropy does not match the change in the modular Hamiltonian’s expectation value, violating the entanglement first law. 
\begin{align}
    \Delta S_{\rm real}+\Delta S_{\rm imaginary}\neq \Delta\langle H\rangle.
\end{align}
This is probably because of the non-smooth connection between the spacelike and timelike surfaces, which leads to an ill-defined variational formula for the area. We note a parallel in the context of computing the first law in dS$_3$ \cite{Fujiki:2025rtx}, where the expected holographic pseudo entropy was found only by resorting to complex coordinates—a strategy that aligns with the systematic method of \cite{Heller:2024whi, Heller:2025kvp} for timelike entanglement entropy, as well as the methodology and findings presented in this paper.

\section{Vanishing of the coefficient of $p^{2m}p_0^{2q}$} \label{prove the coefficient}
Considering the coefficient of $p^{2m}p_0^{2q},m\neq0$,
for simplicity, we calculate it in spherical coordinates 
and omit the overall factor $\Omega_{d-3}$. The remaining integrand is  
\begin{align}
    &e^{ip_0r\cos{\theta}}\frac{\Gamma\bqty{\frac{d+2}{2}}}{2^{2m}m!\Gamma\bqty{\frac{d+2m+2}{2}}}(R^2-r^2)^m(R^2-r^2\cos^2{\theta})\notag\\
    &+e^{ip_0r\cos{\theta}}\frac{\Gamma\bqty{\frac{d+2}{2}}}{2^{2m-2}(m-1)!\Gamma\bqty{\frac{d+2m}{2}}}(R^2-r^2)^{m-1}p_0^{-2}\left(-\frac{r^2\sin^2{\theta}}{d-2}+r^2\cos^2{\theta}\right)\notag\\
    &\propto e^{ip_0r\cos{\theta}}\frac{1}{2m(d+2m)}(R^2-r^2)^m(R^2-r^2\cos^2{\theta})\notag\\
    &+e^{ip_0r\cos{\theta}}(R^2-r^2)^{m-1}p_0^{-2}\left(-\frac{r^2\sin^2{\theta}}{d-2}+r^2\cos^2{\theta}\right).
\end{align}
Then expanding $e^{ip_0r\cos{\theta}}$, the integrand becomes
\begin{align}
    &\sum_{k=0}^{\infty}\frac{(ir)^{2k}}{(2k)!}\frac{1}{2m(d+2m)}(R^2-r^2)^mp_0^{2k}\cos^{2k}{\theta}(R^2-r^2\cos^2{\theta)}\notag\\
    &+\sum_{l=0}^{\infty}\frac{(ir)^{2l}}{(2l)!}(R^2-r^2)^{m-1}p_0^{2l-2}\cos^{2l}{\theta}\left(-\frac{r^2\sin^2{\theta}}{d-2}+r^2\cos^2{\theta}\right).
\end{align}
Here, we ignore that $l\text{ or }k$ is odd as a consequence of the property of the trigonometric function, which causes the integration related to angle turning to zero. Considering the coefficient of $p_0^{2q}$, for $q=-1$, when we integrate out the angle part, it's obvious that the coefficient is zero. For $q\neq -1$, integrating out the angle part , the expression of the integrand is
\begin{align}
    &\frac{1}{2m(d+2m)}(R^2-r^2)^m\cos^{2q}{\theta}(R^2-r^2\cos^2{\theta)}\notag\\
    &+\frac{(ir)^2}{(2q+2)(2q+1)}(R^2-r^2)^{m-1}\cos^{2q+2}{\theta}\left(-\frac{r^2\sin^2{\theta}}{d-2}+r^2\cos^2{\theta}\right),\label{p^2q}
\end{align}
with an overall factor $\frac{(ir)^{2l'}}{(2l')!}$ omitted, where $l'=k \text{ or } l$.  

By using 
 \begin{align}
    \int_0^{\pi}\sin^a{\theta}\cos^b{\theta}d\theta=\frac{\Gamma\bqty{\frac{a+1}{2}}\Gamma\bqty{\frac{b+1}{2}}}{\Gamma\bqty{\frac{a+b+2}{2}}}, b \text{ is even},
 \end{align}
the integration of angle becomes,
\begin{align}
    &\int_0^{\pi}d\theta \sin^{d-3}{\theta}\cos^{2q}{\theta}(R^2-r^2\cos^2{\theta})
    =R^2\frac{\Gamma\bqty{\frac{d-2}{2}}\Gamma\bqty{\frac{2q+1}{2}}}{\Gamma\bqty{\frac{d-1+2q}{2}}}-r^2\frac{\Gamma\bqty{\frac{d-2}{2}}\Gamma\bqty{\frac{2q+1}{2}}(2q+1)}{\Gamma\bqty{\frac{d-1+2q}{2}}(d-1+2q)},\\
    &\int_0^{\pi}d\theta \sin^{d-3}{\theta}\cos^{2q+2}{\theta}\left(-\frac{r^2\sin^2{\theta}}{d-2}+r^2\cos^2{\theta}\right)
    =r^2\frac{\Gamma\bqty{\frac{d-2}{2}}\Gamma\bqty{\frac{2q+1}{2}}}{\Gamma\bqty{\frac{d-1+2q}{2}}}\frac{2(2q+1)(q+1)}{(d+1+2q)(d-1+2q)}.
\end{align}
Then eq.(\ref{p^2q}) can be reformulated as follow with an overall constant factor omitted,
\begin{align}
    \frac{1}{2m(d+2m)}(R^2-r^2)^m\pqty{R^2-r^2\frac{2q+1}{d-1+2q}}-\frac{r^4}{2q+2}(R^2-r^2)^{m-1}\frac{2(q+1)}{(d+1+2q)(d-1+2q)}.
\end{align}
Then integrate over $r$, the coefficient of $p^{2m}p_0^{2q}$ is
\begin{align}
    &\int_0^{R}dr \frac{1}{2m(d+2m)}r^{d-2+2q}(R^2-r^2)^m\pqty{R^2-r^2\frac{2q+1}{d-1+2q}}\notag\\
    &-\int_0^{R}dr\frac{1}{2q+2} r^{d+2+2q}(R^2-r^2)^{m-1}\frac{2(q+1)}{(d+1+2q)(d-1+2q)}\notag\\
    &\propto\frac{1}{2m(d+2m)}\pqty{\frac{m\Gamma\bqty{m}\Gamma\bqty{\frac{d+2q-1}{2}}}{2\Gamma\bqty{\frac{1+d+2m+2q}{2}}}-\frac{2q+1}{d-1+2q}\frac{m\Gamma\bqty{m}\Gamma\bqty{\frac{d+2q-1}{2}}\frac{d+2q-1}{2}}{2\Gamma\bqty{\frac{1+d+2m+2q}{2}}\frac{d+1+2m+2q}{2}}}\notag\\
    &-\frac{2(q+1)}{(d+1+2q)(d-1+2q)(2q+2)}\frac{\Gamma\bqty{m}\Gamma\bqty{\frac{d+2q-1}{2}}\frac{d+2q+1}{2}\frac{d+2q-1}{2}}{2\Gamma\bqty{\frac{1+d+2m+2q}{2}}\frac{1+d+2m+2q}{2}}=0.
\end{align}
Thus, the coefficients of $p^{2m}p_0^{2q},m\neq0$ vanish.

\bibliographystyle{JHEP}
\bibliography{reference}
\end{document}